# Mechanistic transition between inter- and trans-granular creep cracking via a unified crystal plasticity-fracture framework

Weichen Kong [a], Yanwei Dai [b,1], Yue Wang [c], Haitao Wang [c], Yinghua Liu [a,1]

[a] Department of Engineering Mechanics, AML, Tsinghua University, Beijing 100084, China

[b] Institute of Electronics Packaging Technology and Reliability, Department of Mechanics, Beijing University of Technology, Beijing 100124, China

[c] Institute of Nuclear and New Energy Technology, Tsinghua University, Beijing, 100084, China

[1] Corresponding authors:

E-mail addresses: ywdai@bjut.edu.cn (Yanwei Dai), yhliu@mail.tsinghua.edu.cn (Yinghua Liu).

## Abstract

High-temperature creep in metals remains challenging due to the complex interplay of deformation and failure mechanisms across multiple length scales. The trade-off between inter- and trans-granular cracking governs the transition of creep rupture behaviour in metallic materials. However, the micromechanical basis of the transition and its linkage to the macroscopic brittle-to-ductile transition in creep rupture behaviour remain unresolved. Here, the stress-dependent competition between grain-boundary cavitation and slip-induced intragranular cracking is revealed via a unified crystal plasticity-fracture framework. The framework incorporates crystal plasticity, grain-boundary cavitation, and crystallographic crack propagation within an explicit polycrystalline microstructure. The microscale computation reveals the transition from trans-granular-dominated cracking at high stress levels to mixed inter- and trans-granular cracking at intermediate stresses and inter-granular-dominated cracking at lower stresses, accompanied by distinct damage localization modes and the evolution of crack paths from grain boundary to grain-interior propagation. The results further quantify the linkage between microscopic fracture mechanisms and macroscopic creep rupture behaviour, including changes in rupture ductility and creep life. Under high-stress conditions, deformation-driven trans-granular damage localization promotes ductile rupture through extensive strain accumulation, whereas under lower-stress conditions, time-dependent grain-boundary damage accumulation leads to quasi-brittle rupture through localized inter-granular fracture. The study further extends the analysis to representative multiaxial stress states, providing insights into micromechanical creep fracture behaviour and enabling creep life assessment under complex loading conditions. This work provides a framework for further exploring creep fracture involving multiple interacting deformation and damage processes.

# 1. Introduction

At elevated temperatures, creep deformation and fracture are fundamental to the integrity assessment of metallic components operating under severe thermal environments. Several decades ago, Ashby and co-workers showed that creep fracture is governed by different mechanisms depending on stress and temperature, and systematically established fracture-mechanism maps for a wide range of metals and alloys (Ashby et al., 1979; Cocks and Ashby, 1982). Since then, clarifying and quantifying the roles of the underlying micromechanical processes in creep deformation and failure remains an active research topic and a longstanding challenge (Ashraf et al., 2026; Gan et al., 2026; Lei et al., 2025; Liang et al., 2024; Sagara et al., 2026; Sanders et al., 2017).

The creep behaviour of metals varies considerably with stress and temperature, reflecting the coupled operation and transition of multiple deformation and damage mechanisms. These processes involve dislocation glide and climb, diffusion, grain-boundary (GB) sliding, cavity nucleation and growth, precipitation, and microstructural evolution (Raj and Ashby, 1971; Sagara et al., 2026; Zhang et al., 2026). Both experimental investigations and mechanical modelling have revealed the complexity of the mechanisms governing creep deformation and fracture from different perspectives. Experiments by Wang et al. (2021) on a Ni-based superalloy systematically demonstrated shifts in the dominant mechanisms under different temperatures and applied loads, including recrystallization, typical creep cavitation, and material coarsening and embrittlement associated with carbide precipitation. Such complex transitions between creep mechanisms have also been observed in 316 stainless steel (Petkov and Juan, 2024). These observations indicate that the dominant deformation and damage pathways may shift substantially with loading conditions.

Depending on the stress level and loading history, the fracture morphology may evolve from predominantly inter-granular fracture to mixed inter- and trans-granular cracking, and eventually to predominantly trans-granular fracture (Ashby et al., 1979; Cheng et al., 2024; Li et al., 2023). As shown in Fig. 1, the constant-stress creep experiments on Inconel 617 conducted by Wang et al. (2022) reveal a clear transition from predominantly inter-granular to predominantly trans-granular fracture with variations in stress and temperature. At relatively low stress levels, the fracture surfaces exhibit a rock-like morphology characteristic of predominantly inter-granular fracture, with cracks developing primarily along grain boundaries and predominantly perpendicular to the loading direction. In contrast, at relatively high stress levels, numerous intragranular dimples are observed on the fracture surfaces, indicating predominantly trans-granular fracture, accompanied by slip-mediated microscopic voids distributed predominantly parallel to the loading direction. Within the transition regime, both inter-granular and trans-granular damage features coexist, resulting in a mixed fracture morphology. Similar stress-dependent variations in creep fracture morphology have also been widely reported for metallic materials, including 316 stainless steels (Monteiro et al., 2017). Clarifying the dominant regimes and stress-dependent evolution of micromechanical creep cracking and damage is essential for understanding creep fracture behaviour. However, at the microscale, the mechanisms governing this transition and their quantitative representation within micromechanical modelling frameworks remain unclear.

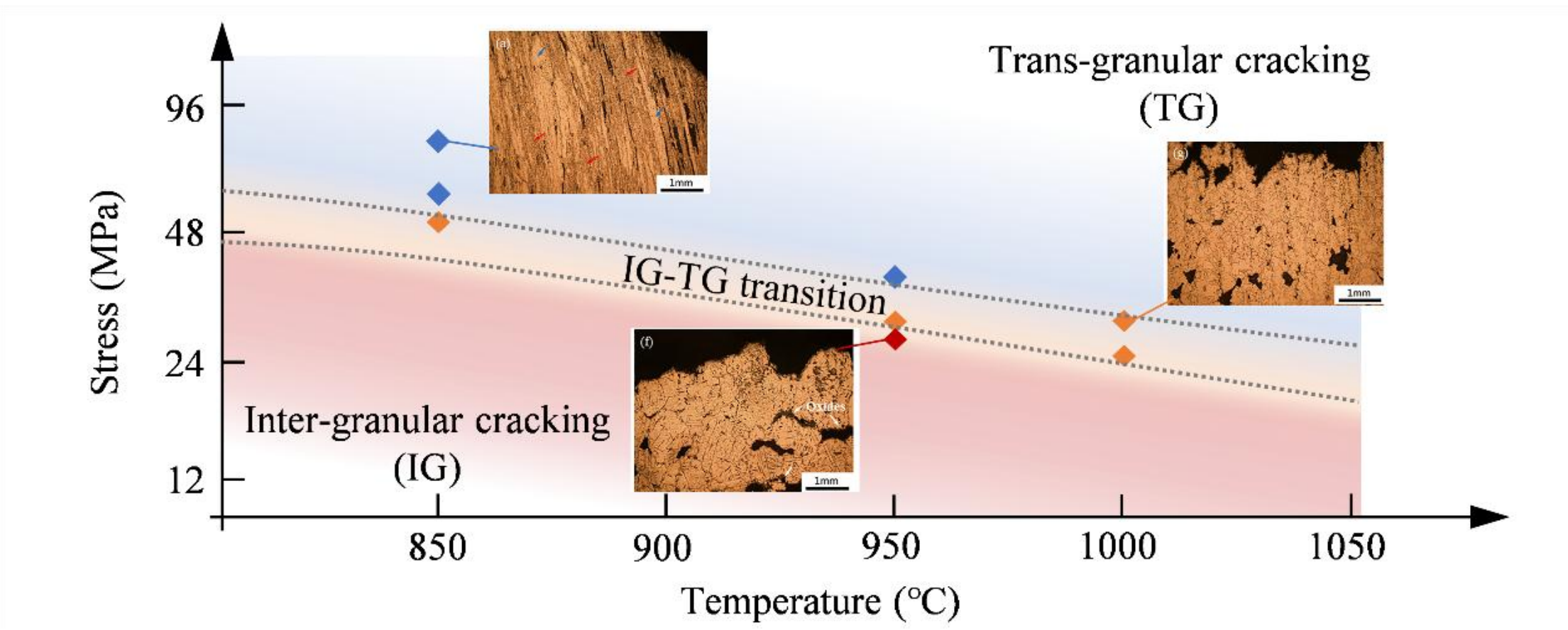

Fig. 1 Creep mechanism map showing the transition from inter-granular to trans-granular fracture in Inconel 617 under different temperatures and stresses

Furthermore, this transition in the underlying mechanisms is directly reflected in the macroscopic creep response and creep failure behaviour of the material. On the one hand, the evolution from predominantly inter-granular to trans-granular creep cracking affects the macroscopic creep behaviour of materials (Wen et al., 2016). In general, metals exhibit relatively high creep rupture ductility at high stress levels, whereas at low stress levels, they tend to exhibit quasi-brittle failure behaviour accompanied by low ductility. On the other hand, in regions with pronounced stress gradients, such as the vicinity of a creep crack tip (Davies et al., 2007), the stress dependence of creep fracture becomes considerably more complex. Dean and Gladwin (2007) demonstrated that variations in applied loading can lead to distinct long-term creep crack-growth behaviour, with the dominant failure mode transitioning between ductile and brittle crack growth. However, the relationship between the macroscopic transition from ductile to brittle creep failure and the corresponding evolution of the underlying micromechanical fracture mechanisms remains incompletely understood.

These observations demonstrate that creep fracture morphology is not solely determined by the occurrence of individual damage mechanisms (Lei et al., 2025; Liang et al., 2024; Salvini et al., 2024), but rather by the stress-dependent evolution and competition of different microscopic damage pathways. To elucidate the grain-scale mechanisms governing creep deformation and fracture, microstructure-sensitive modelling approaches have been extensively developed. Crystal plasticity finite element (CPFE) methods, combined with advanced experimental characterization, have enabled the resolution of local deformation heterogeneity and damage initiation in polycrystalline materials. Ashraf et al. (2024) integrated three-dimensional X-ray diffraction (3DXRD) with CPFE simulations to reveal the effects of crystallographic orientation and local grain neighbourhood on grain-scale creep responses, and further identified plastic deviatoric strain and loading-direction normal stress as important factors associated with GB cavity initiation (Ashraf et al., 2026). Similarly, Shang et al. (2026) combined bicrystal creep experiments with CPFE modelling to investigate the influence of crystallographic orientation, local stress state, and slip activity on cavity nucleation and early-stage growth. These studies demonstrate the capability of CPFE in capturing local deformation evolution and stress redistribution during creep.

In parallel, considerable efforts have been devoted to modelling inter-granular creep damage, which is a dominant failure mechanism under long-term high-temperature conditions. Classical theories established the fundamental understanding of GB cavity nucleation, growth, and coalescence (Cocks and Ashby, 1982; Needleman and Rice, 1980; Tvergaard, 1984; Van Der Giessen et al., 1995), while subsequent cohesive zone and phase-field approaches enabled explicit descriptions of GB degradation and creep crack evolution (Kong et al., 2025; Phan et al., 2017; Zhang et al., 2020). More recently, Gan et al. (2026) coupled crystal plasticity with phase-field damage modelling to investigate inter-granular creep damage in HR3C stainless steel, revealing the relationship between GB cavity nucleation, local energy accumulation, and precipitation-related microstructural degradation. These developments have significantly advanced the understanding of grain-scale deformation and inter-granular damage evolution.

Despite these advances, quantitatively elucidating the stress-dependent transition between inter-granular and trans-granular creep fracture remains challenging within existing modelling frameworks. Current approaches have provided valuable insights into microstructure-sensitive damage evolution, but they generally focus on individual deformation or damage mechanisms rather than their coupled evolution during fracture transition. For example, Li et al. (2025) coupled crystal plasticity with continuum damage mechanics to predict microstructure-dependent creep damage evolution, enabling the characterization of damage accumulation and spatial distribution within polycrystalline structures. However, such continuum-based approaches still face limitations in explicitly resolving local crack initiation and propagation paths at the grain scale. Alternatively, phase-field methods have demonstrated significant capability in describing complex crack evolution, but when crystallographic information or GB physics are not explicitly incorporated, they cannot fully distinguish the different micromechanical origins of cracking associated with grain boundaries and grain interiors (Si et al., 2025). Moreover, existing micromechanical fracture models typically treat inter-granular and trans-granular cracking separately, with cohesive zone models (CZM) mainly employed for GB degradation (Phan et al., 2017) and extended finite element methods (XFEM) used to describe crack propagation within grain interiors (Wilson and Dunne, 2019). Consequently, a unified framework that simultaneously captures these two microstructure-sensitive cracking processes and their evolution within explicit polycrystalline structures remains limited.

Despite significant progress in understanding creep deformation and inter-granular damage evolution, several fundamental questions regarding stress-dependent fracture transition remain unresolved. First, although transitions from inter-granular to trans-granular creep fracture have been widely reported, the micromechanical origin controlling this transition remains unclear. In particular, how GB degradation and deformation-driven intragranular cracking evolve with stress and compete for dominance has not been systematically characterized. Second, the quantitative relationship between microscopic fracture mechanism evolution and macroscopic creep behaviour, including rupture ductility and creep life, remains insufficiently understood. It remains unclear how changes in crack morphology and damage pathways influence the overall failure response of metallic materials. Third, existing modelling approaches are generally developed to describe individual fracture mechanisms rather than their coupled evolution. While CZM and phase-field approaches provide effective descriptions of GB damage, and XFEM-based approaches enable intragranular crack propagation, a unified grain-scale framework capable of simultaneously capturing GB degradation, crystallographically driven intragranular cracking, and their transition within explicit polycrystalline structures is still lacking.

To address these issues, this study establishes a mechanistic understanding of the stress-dependent transition between inter-granular and trans-granular creep fracture by clarifying the competition and interaction between time-dependent GB degradation and deformation-driven intragranular cracking. Unlike existing approaches that treat these fracture mechanisms separately, a unified crystal plasticity-fracture framework is developed to simultaneously capture grain-scale deformation, GB cavitation-induced damage, and crystallographically controlled trans-granular crack propagation within explicit polycrystalline structures. Through systematic comparison with experimental observations, the proposed framework elucidates how the evolving dominance of different fracture pathways leads to distinct damage localization modes, governs the transition from inter-granular to mixed and trans-granular fracture morphologies, and correlates with the associated variation in creep rupture ductility and creep life. Furthermore, the framework is extended to multiaxial loading conditions to investigate the influence of complex stress states on the evolution of distinct damage pathways and their contribution to creep life.

# 2. Stress-sensitive inter- and trans-granular creep cracking in metals

To clarify the physical problems addressed in this work, this section discusses the stress-sensitive microscale creep cracking behaviour and its influence on macroscopic creep properties, with particular emphasis on the evolution from inter-granular to trans-granular creep cracking and from quasi-brittle to ductile rupture.

## 2.1. Effect of stress on the creep cracking behaviour

Under constant-stress creep loading, the creep cracking behaviour of metals at the microscale is highly sensitive to the applied stress level, as widely observed in various materials. As shown in Fig. 2a, the material exhibits predominantly trans-granular creep cracking under relatively high stress levels. The fracture surface is characterized by small dimples with sizes smaller than the average grain size, resulting from intra-granular dislocation accumulation, crack initiation, and subsequent void growth within grains. In contrast, as shown in Fig. 2b, under relatively low stress conditions, the microscale creep cracking behaviour exhibits distinct characteristics, with inter-granular creep cracking dominated by GB failure. The fracture surface exhibits a rough rock-candy morphology, which is characteristic of inter-granular cracking. Similar mechanisms have been observed experimentally across a wide range of materials, including nickel-based superalloys and stainless steels, as exemplified by Alloy 617 (Wang et al., 2022) and 316 stainless steel (Monteiro et al., 2017), suggesting that such stress-sensitive creep cracking behaviour is a general feature of high-temperature structural materials. Although these phenomena and associated mechanisms have been identified experimentally and summarized empirically, quantitative multiscale modelling remains challenging. Existing approaches have not fully incorporated the micromechanical behaviour of polycrystalline metals to describe the stress-sensitive evolution of inter-granular and trans-granular creep cracking.

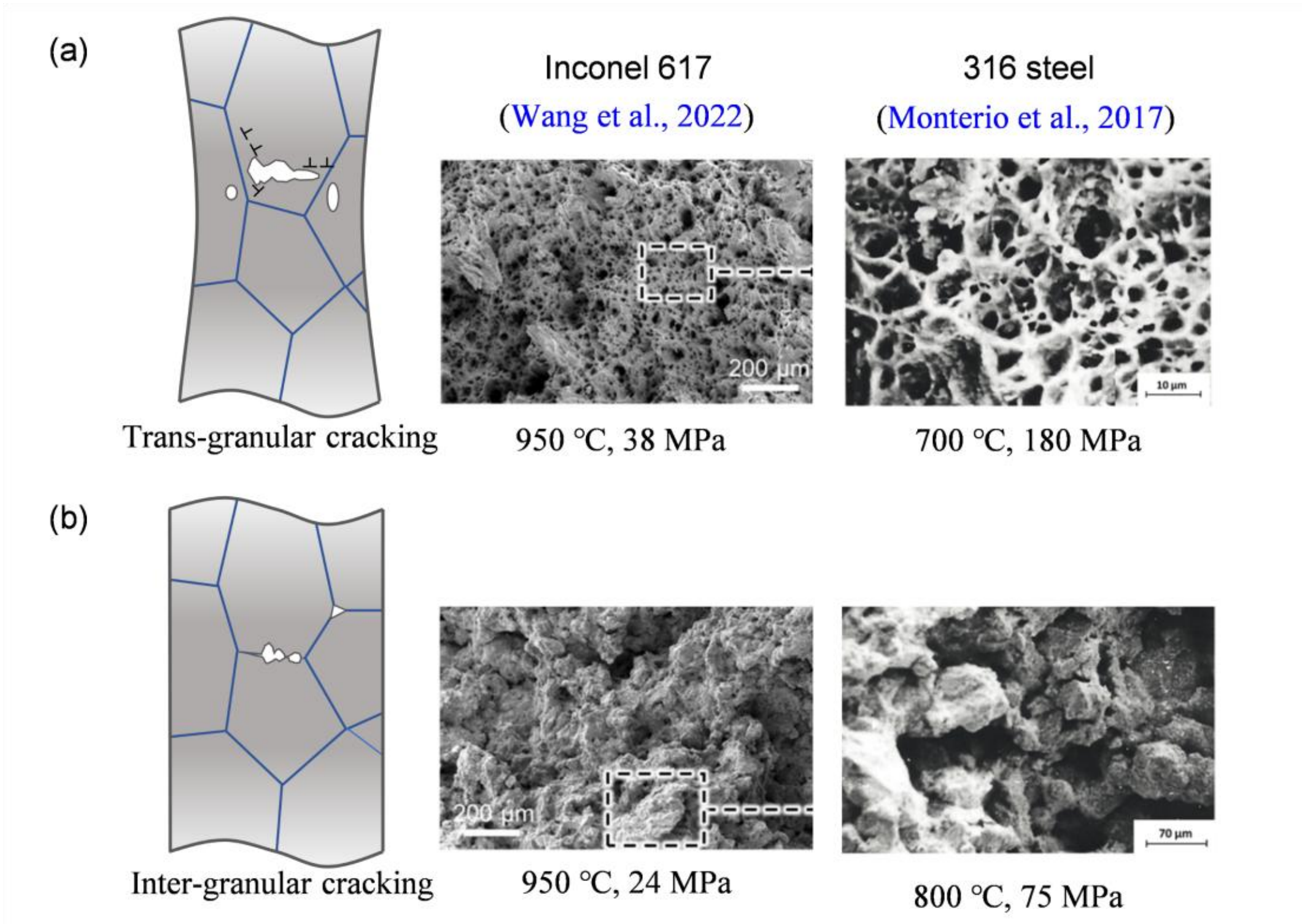

Fig. 2 Schematic illustration of stress-sensitive creep cracking modes under different temperature and stress conditions: (a) predominantly trans-granular creep cracking at higher stress levels; (b) predominantly inter-granular creep cracking at lower stress levels.

Previous studies on constant-stress creep have largely focused on inter-granular creep cracking under relatively low stress conditions. Considerable efforts have been devoted to distinguishing different GB cavity nucleation and growth mechanisms (Cocks and Ashby, 1982; Van Der Giessen et al., 1995), as well as their contributions to GB damage and failure (Zhang et al., 2020). In contrast, the present work focuses on a broader stress range covering both relatively low and high stress levels, where inter-granular, trans-granular and mixed creep cracking can all be observed. Such stress conditions are commonly encountered in stress concentration regions of high-temperature components and in the vicinity of creep crack tips.

## 2.2. Transition from quasi-brittle to ductile rupture of creeping metals

Furthermore, the differences in microscale creep cracking modes are reflected in the macroscopic creep rupture properties. Extensive experimental observations have shown that the creep rupture ductility under constant-stress creep is strongly dependent on the applied stress level (Wen et al., 2016). As shown in Fig. 3, experimental observations indicate that the creep rupture ductility, i.e., the creep rupture strain, increases with increasing applied stress and gradually approaches a plateau. At relatively low stress levels, the material exhibits low creep rupture ductility. In this regime, inter-granular cracking is dominant, with GB cavity nucleation and growth governed by diffusion and bulk creep processes. Consequently, the material exhibits an overall quasi-brittle rupture (see Fig. 3a). With the increasing applied stress, a mixed creep cracking mode involving both inter-granular creep cracking associated with GB cavities and trans-granular void cracking associated with plastic deformation develops (see Fig. 3b). In this regime, the macroscopic creep rupture ductility increases. At higher stress levels, trans-granular plastic deformation-induced void nucleation and growth become the dominant processes during creep failure (see Fig. 3c). Consequently, the material exhibits a ductile rupture at the macroscale.

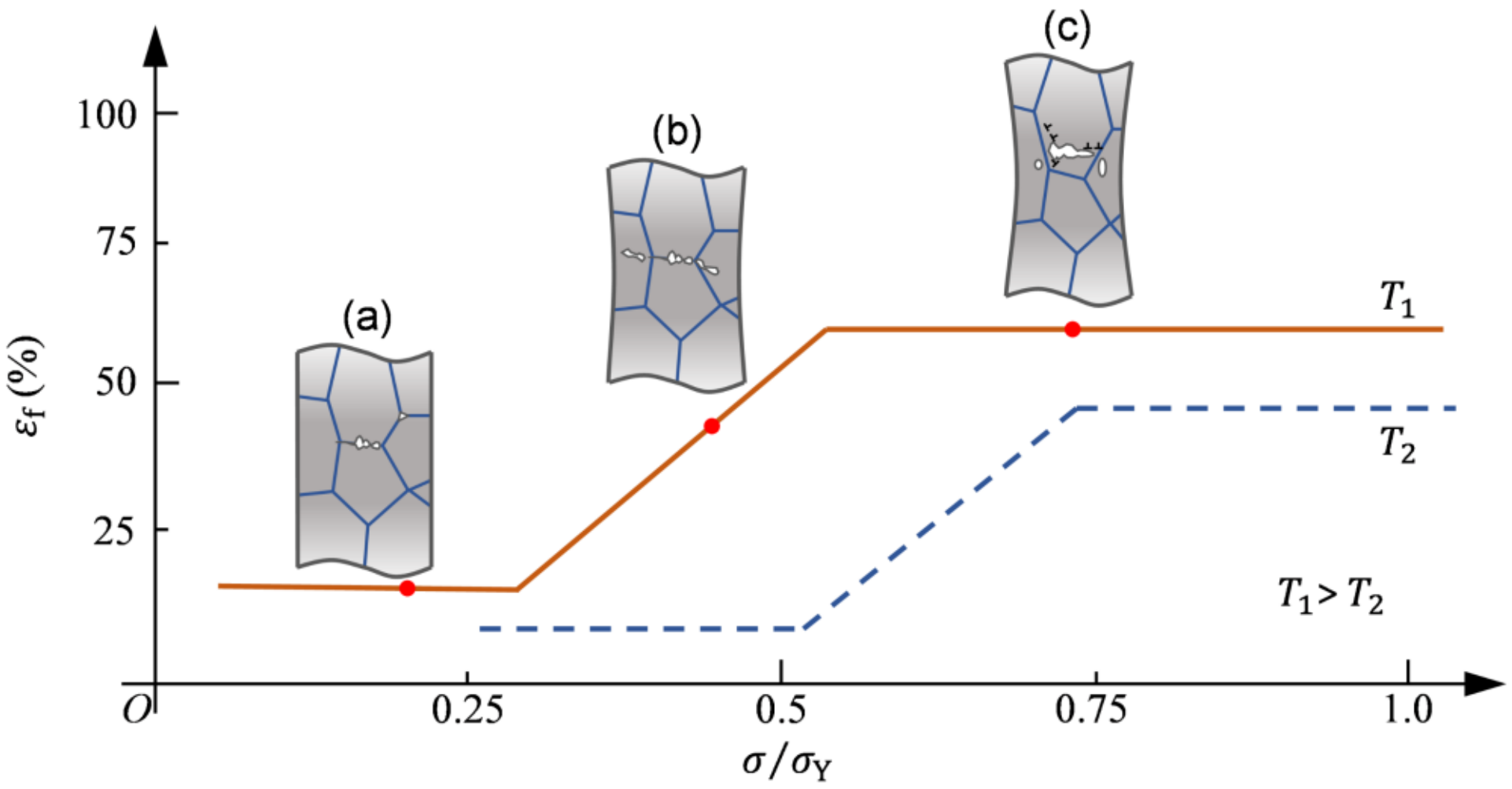

Fig. 3 Schematic illustrations of the influence of microscale creep cracking modes on the creep rupture ductility under different stress levels: (a) predominantly inter-granular creep cracking; (b) mixed inter- and trans-granular creep cracking; (c) predominantly trans-granular creep cracking.

The transition from quasi-brittle to ductile rupture in creeping metals exhibits similar characteristics across different temperature conditions. The dependence of macroscopic creep rupture ductility on microscale creep cracking behaviour plays an important role in the safety assessment of high-temperature structures. Therefore, quantitative modelling of creep damage processes and the establishment of a link between microscale cracking behaviour and macroscopic creep properties are essential for predicting creep failure.

# 3. A unified crystal plasticity-fracture framework for inter- and trans-granular creep cracking

In this work, a unified crystal plasticity-fracture modelling framework is developed to simulate deformation, inter- and trans-granular creep cracking in polycrystalline metals. The crystal plasticity model is employed to describe grain-level plastic and creep deformation, while the cohesive zone model (CZM) is used to capture GB damage. In addition, the combination of the CZM and extended finite element method (XFEM) is adopted to describe the initiation and propagation of trans-granular cracks. Considering the requirements of the microscale modelling framework and the availability of well-established experimental data, the nickel-based superalloy Inconel 617 is selected as the representative material in this work.

## 3.1. Crystal plasticity finite element model for creep alloys

The crystal plasticity finite element (CPFE) method is employed to describe the plastic and creep deformation of the grains. A CPFE model incorporating both crystallographic slip and dislocation climb mechanisms is adopted, which has been successfully applied and validated in previous studies (Kong et al., 2025; Phan et al., 2017; Zhang and Oskay, 2016). The kinematic framework of the crystal plasticity model follows the formulation proposed by Marin and Dawson (1998). The deformation gradient $\mathbf{F}$ is expressed as:

$$\mathbf{F} = \mathbf{V}^{\mathrm{e}} \cdot \mathbf{R}^{\mathrm{e}} \cdot \mathbf{F}^{\mathrm{p}} \tag{1}$$

in which $\mathbf{F}^{\mathrm{p}}$, $\mathbf{R}^{\mathrm{e}}$, $\mathbf{V}^{\mathrm{e}}$ denote plastic deformation gradient, elastic rotation, and the left symmetric elastic stretch tensor, respectively. This multiplicative decomposition introduces four configurations: the reference configuration $\mathbb{C}_0$, the intermediate configuration $\bar{\mathbb{C}}$ associated with the plastic deformation gradient, the intermediate configuration $\tilde{\mathbb{C}}$ after the elastic rotation, and the current configuration $\mathbb{C}$ obtained after the elastic stretch. The evolution of constitutive variables is evaluated in the configuration $\tilde{\mathbb{C}}$. The velocity gradient $\tilde{\mathbf{L}}^{\mathrm{p}}$ is then decomposed into the rate of deformation tensor $\tilde{\mathbf{D}}^{\mathrm{p}}$ and the spin tensor $\tilde{\mathbf{W}}^{\mathrm{p}}$:

$$\tilde{\mathbf{L}}^{\mathrm{p}} = \tilde{\mathbf{D}}^{\mathrm{p}} + \tilde{\mathbf{W}}^{\mathrm{p}} \tag{2}$$

Subsequently, the rate of deformation and spin tensors are expressed as the sum of contributions from dislocation glide and climb mechanisms (Phan et al., 2017).

$$\tilde{\mathbf{D}}^{\mathrm{p}} = \tilde{\mathbf{D}}^{\mathrm{P}}_{\mathrm{g}} + \tilde{\mathbf{D}}^{\mathrm{P}}_{\mathrm{c}} \tag{3}$$

$$\tilde{\mathbf{W}}^{\mathrm{p}} = \tilde{\mathbf{W}}^{\mathrm{P}}_{\mathrm{g}} + \tilde{\mathbf{W}}^{\mathrm{P}}_{\mathrm{c}} \tag{4}$$

The subscripts ‘g’ and ‘c’ represent the contributions from dislocation glide and climb mechanisms, respectively. Specifically, these terms are given as:

$$\tilde{\mathbf{D}}^{\mathrm{P}}_{\mathrm{g}} = \sum_{\alpha=1}^{n} \dot{\gamma}^{\alpha}_{\mathrm{g}} \left( \tilde{\mathbf{s}}^{\alpha} \otimes \tilde{\mathbf{m}}^{\alpha} \right)_{\mathrm{S}} \tag{5}$$

$$\tilde{\mathbf{D}}^{\mathrm{P}}_{\mathrm{c}} = \sum_{\alpha=1}^{n} \dot{\gamma}^{\alpha}_{\mathrm{c}} \left( \tilde{\mathbf{K}}^{\alpha} \right)_{\mathrm{S}} \tag{6}$$

$$\tilde{\mathbf{W}}^{\mathrm{P}}_{\mathrm{g}} = \dot{\mathbf{R}}^{\mathrm{e}} \cdot \left( \mathbf{R}^{\mathrm{e}} \right)^{\mathrm{T}} + \sum_{\alpha=1}^{n} \dot{\gamma}^{\alpha}_{\mathrm{g}} \left( \tilde{\mathbf{s}}^{\alpha} \otimes \tilde{\mathbf{m}}^{\alpha} \right)_{\mathrm{A}} \tag{7}$$

$$\tilde{\mathbf{W}}^{\mathrm{P}}_{\mathrm{c}} = \sum_{\alpha=1}^{n} \dot{\gamma}^{\alpha}_{\mathrm{c}} \left( \tilde{\mathbf{K}}^{\alpha} \right)_{\mathrm{A}} \tag{8}$$

in which $\tilde{\mathbf{K}}^{\alpha} = \tilde{\mathbf{s}}^{\alpha} \otimes \tilde{\boldsymbol{\chi}}^{\alpha}$ and $\tilde{\boldsymbol{\chi}}^{\alpha} = \tilde{\mathbf{m}}^{\alpha} \times \tilde{\mathbf{t}}^{\alpha}$. $\tilde{\mathbf{s}}^{\alpha}$, $\tilde{\mathbf{m}}^{\alpha}$ and $\tilde{\mathbf{t}}^{\alpha}$ are the $\alpha$-slip direction, vector normal to the $\alpha$-slip plane and the tangent to the dislocation line. $(\bullet)_{\mathrm{S}}$ and $(\bullet)_{\mathrm{A}}$ represent the symmetric and skew-symmetric parts of the tensor. For each slip system, the shear rate is given by:

$$\dot{\gamma}_{\mathrm{g}}^{\alpha}=\dot{\gamma}_{0\mathrm{g}}\exp\left\{-\frac{F_0}{k\theta}\left\langle 1-\left\langle\frac{\left|\tau^{\alpha}-B^{\alpha}\right|-S^{\alpha}\mu/\mu_0}{\hat{\tau}_0\mu/\mu_0}\right\rangle^{p_g}\right\rangle^{q_g}\right\}\mathrm{sgn}\left(\tau^{\alpha}-B^{\alpha}\right) \tag{9}$$

with resolved shear stress $\tau^{\alpha}=\boldsymbol{\tau}:\left(\tilde{\mathbf{s}}^{\alpha}\otimes\tilde{\mathbf{m}}^{\alpha}\right)_{\mathrm{S}}$. $\dot{\gamma}_0$ is the reference shear strain rate, $k$ is the Boltzmann constant, $\theta$ is the absolute temperature, and $F_0$ is the activation energy. The parameter $\hat{\tau}_0$ specifies the threshold stress for thermally activated dislocation motion. The shear modulus at the current temperature and its corresponding value at 0 K are denoted by $\mu$ and $\mu_0$, respectively. The quantities $S^{\alpha}$ and $B^{\alpha}$ denote the slip resistance and back stress of $\alpha$ -slip system, respectively. The dislocation climb-influenced shear rate is determined according to Phan et al. (2017):

$$\dot{\gamma}_{\mathrm{c}}^{\alpha}=\dot{\gamma}_{0\mathrm{c}}\exp\left(-\frac{F_0}{k\theta}\right)\left(\frac{\left|\tau_{\mathrm{c}}^{\alpha}-B_{\mathrm{c}}^{\alpha}\right|}{\hat{\tau}_{0\mathrm{c}}}\right)^{p_{\mathrm{c}}}\mathrm{sgn}\left(\tau_{\mathrm{c}}^{\alpha}-B_{\mathrm{c}}^{\alpha}\right) \tag{10}$$

where $\hat{\tau}_{0\mathrm{c}}$ and $p_c$ represent the threshold creep stress and the creep exponent, respectively. The resolved shear stress associated with the climb system is defined as $\tau_{\mathrm{c}}^{\alpha}=\boldsymbol{\tau}:\left(\tilde{\mathbf{K}}^{\alpha}\right)_{\mathrm{dev}}$ . represents the back stress arising from climb-induced shear deformation. The above CPFE framework and constitutive formulations have been previously developed and validated by Phan et al. (2017) and Zhang and Oskay (2016).

The slip resistance $S^{\alpha}$ in Eqn. (9) is expressed as:

$$\dot{S}^{\alpha}=\left[h_{\mathrm{S}}-d_{\mathrm{D}}\left(S^{\alpha}-\bar{S}^{\alpha}\right)\right]\left|\dot{\gamma}^{\alpha}\right|-h_2\left(S^{\alpha}-S_0^{\alpha}\right)H\left(\dot{\gamma}_{\mathrm{th}}-\sum_{\alpha=1}^{n}\left|\dot{\gamma}^{\alpha}\right|\right) \tag{11}$$

where $h_{\mathrm{S}}$ and $d_{\mathrm{D}}$ are the hardening and dynamic recovery coefficients, respectively. The parameter $\bar{S}^{\alpha}$ denotes the steady-state slip resistance of $\alpha$ -slip system, while $h_2$ characterizes the static recovery process. The quantities $S_0^{\alpha}$ and $\dot{\gamma}_{\mathrm{th}}$ represent the initial slip resistance and the threshold shear rate for static recovery, respectively. The Heaviside function is denoted by $H(\bullet)$. Following the formulation proposed by Lin et al. (2010), the back stress $B^{\alpha}$ is evaluated as:

$$\dot{B}^{\alpha} = \left[ h_{\mathrm{B}} - D^{\alpha} B^{\alpha} \operatorname{sgn}\left(\dot{\gamma}^{\alpha}\right) \right] \dot{\gamma}^{\alpha} \tag{12}$$

Here, $h_{\mathrm{B}}$ and $D^{\alpha}$ correspond to the hardening and dynamic recovery parameters governing the evolution of the back stress, respectively. The dynamic recovery coefficient $D^{\alpha}$ is defined as:

$$D^{\alpha} = \frac{h_{\mathrm{B}} \mu_0}{S^{\alpha}} \left( \frac{\mu_0'}{f_{\mathrm{co}}} - \mu \right)^{-1} \tag{13}$$

in which $\mu_0'$ is the local slip shear modulus at 0 K, and $f_{\mathrm{co}}$ is a coupling parameter associated with the internal slip variables. The evaluation of the back stress in Eqn. (10) is expressed as (Staroselsky and Cassenti, 2011):

$$\dot{B}_{\mathrm{c}}^{\alpha} = h_{\mathrm{c}} \left( \dot{\gamma}_{\mathrm{c}}^{\alpha} B_{\infty} - \left| \dot{\gamma}_{\mathrm{c}}^{\alpha} \right| B_{\mathrm{c}}^{\alpha} \right) \tag{14}$$

where $B_{\infty}$ represents the saturation value of back stress, and $h_{\mathrm{c}}$ is the material parameter controlling the evolution of back stress.

## 3.2. Inter-granular cohesive fracture with GB cavity nucleation and growth

It is generally recognized that, under constant-stress creep conditions, GB failure is governed by the nucleation and growth of GB cavities. As shown in Fig. 4a, assuming that the GB cavities are spherical caps, the volume $V$ of a single cavity is given by (Onck and van der Giessen, 1999):

$$V = \frac{4}{3} \pi a^3 h\left(\psi_{\mathrm{cav}}\right) \tag{15}$$

in which $a$ is the radius of the cavity, $h\left(\psi_{\mathrm{cav}}\right)$ is a shape function, which is expressed as:

$$h\left(\psi_{\mathrm{cav}}\right) = \frac{\left[ \left(1 + \cos\psi_{\mathrm{cav}}\right)^{-1} - \left(\cos\psi_{\mathrm{cav}}\right)/2 \right]}{\sin\psi_{\mathrm{cav}}} \tag{16}$$

where $\psi_{\mathrm{cav}}$ denotes the dihedral angle shown in Fig. 4a, and is set as constant $\psi_{\mathrm{cav}} = 70°$ for Inconel 617 (Sanders et al., 2017). Hence, the radius $a$ and its evolution are expressed as:

$$a = \left[ \frac{3V}{4\pi h(\psi_{\text{cav}})} \right]^{1/3} \tag{17}$$

$$\dot{a} = \frac{\dot{V}}{4\pi a^2 h(\psi_{\text{cav}})} \tag{18}$$

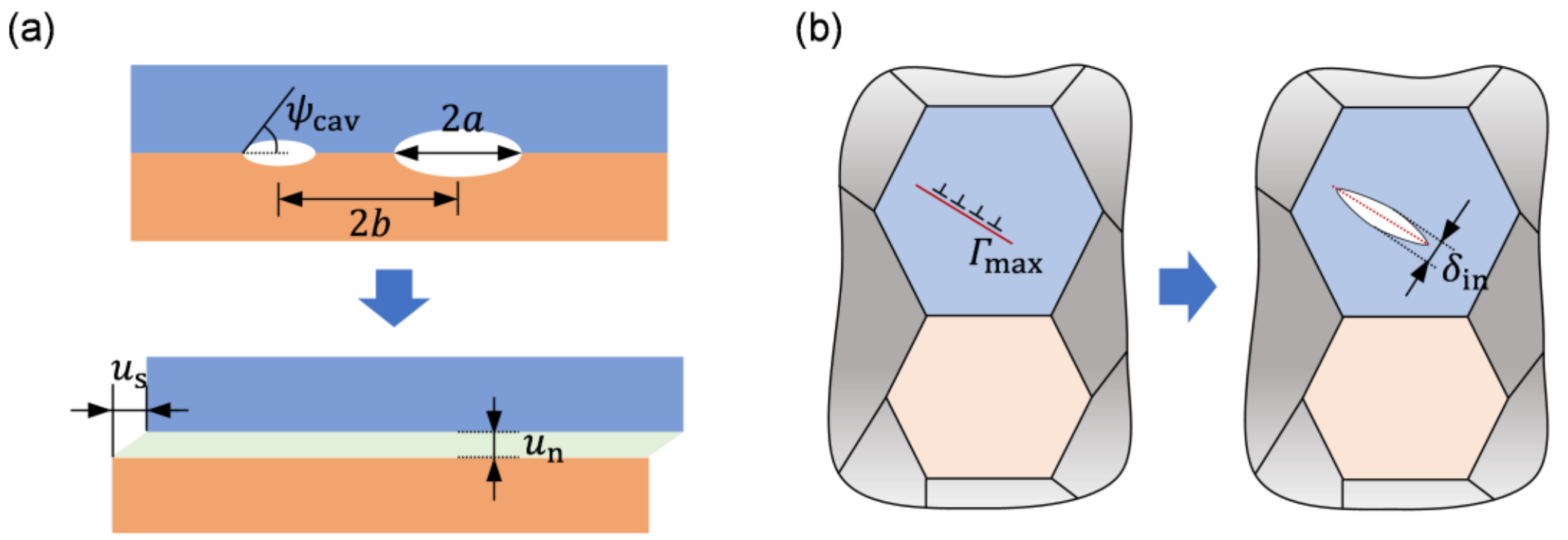

Fig. 4 Schematic illustrations of the micromechanical modelling approaches for (a) GB cavities represented by an equivalent cohesive zone model and (b) trans-granular crack initiation and propagation.

In this study, both nucleation and growth processes of GB cavities under creep conditions are considered. To describe the behaviour of GB cavities within the CPFE framework, a cohesive zone model is adopted in the present study. As shown in Fig. 4a, the normal and tangential displacement separations across the cohesive interface are denoted by $u_{\text{n}}$ and $u_{\text{t}}$, respectively. When accounting for the equivalent damage induced by GB cavities, the traction-separation relations are expressed as:

$$T_{\text{n}} = g(D_{\text{GB}}) k_{\text{n}} \frac{u_{\text{n}}}{\delta_{\text{c}}} \tag{19}$$

$$T_{\text{t}} = g(D_{\text{GB}}) k_{\text{t}} \frac{u_{\text{n}}}{\delta_{\text{c}}} \tag{20}$$

where $k_{\text{n}}$ and $k_{\text{t}}$ are the initial normal and tangential cohesive stiffnesses prior to the onset of equivalent damage, respectively, and $g(D_{\text{GB}})$ is the stiffness degradation function governed by the GB damage variable $D_{\text{GB}}$.

The nucleation process of GB cavities is highly complex (Abe and Tabuchi, 2004;

Zhang et al., 2020), involving microstructural features such as second-phase particles that are difficult to explicitly represent within the CPFE framework. Therefore, phenomenological approaches are commonly adopted to represent the nucleation behaviour of GB cavities. According to Tvergaard's model, the nucleation rate of cavities is assumed to be a function of the equivalent creep strain rate and the normal stress acting on the GB (Tvergaard, 1984). In this work, to directly couple the cavity nucleation behaviour with the GB cohesive zone model, it is assumed that:

$$\dot{N} = A_{\mathrm{N}} \left\langle \frac{T_{\mathrm{n}} - T_{\mathrm{N}}^{\mathrm{c}}}{T_{\mathrm{N}}} \right\rangle^{p_{\mathrm{N}}} \left( \frac{T_{\mathrm{eq}}}{T_{\mathrm{N}}^{0}} \right)^{q_{\mathrm{N}}} \left( \frac{N_{\max} - N}{N_{\max}} \right) \tag{21}$$

where $T_{\mathrm{n}}$ represents the normal tensile traction in the GB cohesive zone model, $T_{\mathrm{N}}^{\mathrm{c}}$ denotes the critical normal traction for cavity nucleation, and $T_{\mathrm{N}}$ and $T_0$ are the reference normal traction and equivalent traction, respectively. $A_{\mathrm{N}}$ is a model constant, while $p_{\mathrm{N}}$ and $q_{\mathrm{N}}$ are the nucleation exponents. The equivalent traction $T_{\mathrm{eq}}$ is defined as:

$$T_{\mathrm{eq}} = \sqrt{\langle T_{\mathrm{n}} \rangle^2 + \frac{(T_{\mathrm{t}})^2}{\alpha}} \tag{22}$$

in which $T_{\mathrm{t}}$ represents the tangential traction in the 2D CZM, and $\alpha$ denotes the weighting factor for the contribution of tangential traction, with $k_{\mathrm{t}} = \alpha k_{\mathrm{n}}$. In Eqn. (21), a saturation mechanism proposed by Nassif et al. (2019) for cavity nucleation is also introduced by assuming that the cavity nucleation density, $N$, is bounded by a maximum value, $N_{\max}$. Eqn. (21) indicates that the evolution rate of $N$ decreases as it approaches the saturation limit $N_{\max}$, and eventually tends to zero. The cavity spacing is then updated according to the cavity nucleation density as:

$$b = b_i \sqrt{\frac{N_i}{N}} \tag{23}$$

in which $b_i$ and $N_i$ are the initial half-spacing between adjacent cavities and the initial cavity nucleation density, respectively.

Furthermore, cavity growth is generally regarded as the dominant mechanism governing GB failure (Cocks and Ashby, 1982; Sham and Needleman, 1983; Van Der Giessen et al., 1995). It typically occurs through two distinct mechanisms: vacancy diffusion-controlled growth and deformation-controlled growth driven by the deformation of adjacent grains. The cavity volume growth is formulated as:

$$\dot{V} = \dot{V}_{\mathrm{D}} + \dot{V}_{\mathrm{C}} \tag{24}$$

in which $\dot{V}_{\mathrm{D}}$ represents the cavity growth induced by diffusion, and $\dot{V}_{\mathrm{C}}$ denotes the cavity growth caused by matrix creep deformation.

In fact, the growth behaviour of GB creep cavities has been extensively studied (Cocks and Ashby, 1982; Needleman and Rice, 1980), leading to a comprehensive understanding of the underlying growth mechanisms. In practice, the growth behaviour of GB cavities can also be effectively described using phenomenological damage models, such as the Kachanov-Rabotnov model, without explicitly considering the underlying cavity growth mechanisms (Phan et al., 2017). To retain the physical significance of the dominant GB damage mechanism while improving the robustness of subsequent finite element simulations, the cavity growth model and its coupling with the damage evolution are simplified phenomenologically in this work.

The diffusion-controlled cavity growth component is formulated based on classical GB cavitation theories, where vacancy diffusion driven by GB opening stress is considered as the dominant mechanism for cavity enlargement (Cocks and Ashby, 1982; Needleman and Rice, 1980),. Therefore, the diffusion-controlled growth term is defined as:

$$\dot{V}_{\mathrm{D}} = A_{\mathrm{D}} \left( \frac{b_{\mathrm{i}}}{b} \right)^{r_{\mathrm{D}}} \left\langle \frac{T_{\mathrm{n}} - T_{\mathrm{c}}}{T_0} \right\rangle^{m_{\mathrm{D}}} \tag{25}$$

in which the normal traction $T_{\mathrm{n}}$ is introduced as the primary driving force, representing the effect of tensile stress on vacancy migration and accumulation at the cavity surface. $A_{\mathrm{D}}$, $r_{\mathrm{D}}$, and $m_{\mathrm{D}}$ are the coefficient and exponents controlling the diffusion-dominated cavity growth rate, respectively. In addition, the cavity spacing $b$ is incorporated to account for the interaction between neighboring cavities. As cavity nucleation proceeds, the reduction in cavity spacing shortens the diffusion distance and enhances vacancy transport, leading to an accelerated growth rate (Hull and Rimmer, 1959). Thus, the normalized spacing ratio $b_{\mathrm{i}}/b$ is introduced as a geometric correction factor to capture the influence of cavity distribution on diffusion-controlled growth.

The creep accommodation-controlled cavity growth component describes cavity enlargement driven by local creep deformation of the surrounding grains. Creep-assisted growth depends on the deformation accommodation capability around the cavity. Therefore, an equivalent traction $T_{\mathrm{eq}}$, incorporating both normal and tangential GB tractions, is adopted as the driving force to represent the coupled effects of cavity opening and GB sliding (Cocks and Ashby, 1982; Needleman and Rice, 1980). Hence, the creep growth term is defined as:

$$\dot{V}_{\mathrm{C}} = A_{\mathrm{C}} \left( \frac{V}{V_{\mathrm{i}}} \right)^{r_{\mathrm{C}}} \left( \frac{T_{\mathrm{eq}}}{T_0} \right)^{m_{\mathrm{C}}} \tag{26}$$

where $A_{\mathrm{C}}$, $r_{\mathrm{C}}$, and $m_{\mathrm{C}}$ are the coefficient and exponents controlling the diffusion-dominated cavity growth rate, respectively. In addition, the normalized cavity volume $V/V_{\mathrm{i}}$ is introduced to capture the accelerated growth of larger cavities, as increasing cavity size leads to stronger local stress concentration and enhanced creep accommodation. Thus, based on Eqns. (17) and (23), the cavity nucleation and growth processes can be related to the evolution of the cavity geometry. Hence, the cavity coalescence leading to GB failure is described using the ratio between the cavity radius and the cavity half-spacing:

$$\chi = \frac{a}{b} = \frac{1}{b} \left[ \frac{3V}{4\pi h(\psi_{\mathrm{cav}})} \right]^{1/3} \tag{27}$$

The GB damage is defined as:

$$D_{\mathrm{GB}} = \frac{\chi - \chi_{\mathrm{i}}}{\chi_{\mathrm{c}} - \chi_{\mathrm{i}}} \tag{28}$$

where $\chi_{\mathrm{i}}$ is the initial coalescence ratio, and $\chi_{\mathrm{c}}$ is the critical coalescence ratio.

Based on the GB damage defined in Eqn. (28), the stiffness degradation function in Eqns. (19) and (20) is defined as:

$$g\left(D_{\mathrm{GB}}\right)=\left(1-D_{\mathrm{GB}}\right)^{2} \tag{29}$$

At the modelling level, the value of $D_{\mathrm{GB}}$ is defined within the range of $0 \leq D_{\mathrm{GB}} < 1$. However, to avoid numerical singularities during the simulations, the upper limit is set as $D_{\mathrm{GB}} < 0.999$. Therefore, the above phenomenological damage model, derived from the underlying physical mechanisms, provides the formulation for describing GB damage evolution and degradation. This formulation enables the evolution of GB cavity damage to be directly coupled with the traction-separation response of the cohesive interface, allowing the initiation and propagation of inter-granular cracks to be captured within the CPFE framework.

## 3.3. Trans-granular crack initiation and propagation

The extended finite element method (XFEM) has been widely employed to describe crack initiation and propagation within polycrystalline materials, particularly for capturing intragranular crack evolution without requiring predefined crack paths (Wilson and Dunne, 2019). To capture both the initiation and propagation of trans-granular cracks under creep conditions, the XFEM is employed in the present study. As shown in Fig. 4b, trans-granular crack initiation is governed by the accumulated slip history variable obtained from the crystal plasticity model, whereas crack propagation is simulated using XFEM coupled with an energy-based cohesive degradation law.

The initiation of trans-granular cracks is governed by an accumulated slip criterion:

$$\Gamma \geq \Gamma_{\mathrm{c}} \tag{30}$$

in which $\Gamma$ represents the accumulated slip.

$$\Gamma=\int_{0}^{t} \dot{\gamma}_{\mathrm{tot}} \mathrm{d} t \tag{31}$$

$$\dot{\gamma}_{\mathrm{tot}}=\sum_{\alpha=1}^{N}\left|\dot{\gamma}_{\alpha}^{\mathrm{c}}\right| \tag{32}$$

in which $\dot{\gamma}_{\mathrm{tot}}$ represents the total slip rate, while $\dot{\gamma}_{\alpha}^{\mathrm{c}}$ denotes the climb-related slip rate associated with the $\alpha$-slip system, and $N$ is the number of the activated slip systems. In Eqn. (32), only the climb-assisted slip process is considered. This assumption is adopted because the present study focuses on the stress regime associated with the transition from trans- to inter-granular creep fracture. Under such conditions, the applied stress remains below the macroscopic yield strength of the material, and long-term deformation is dominated by thermally activated creep mechanisms. Therefore, conventional glide-dominated plasticity is neglected, while climb-assisted slip is retained to describe the creep deformation process.

For the crack propagation direction, the most active slip system is selected among all slip systems, and the activity parameter is defined as follows:

$$A_\alpha = \left|\dot{\gamma}_\alpha^{\mathrm{g}}\right| + \left|\dot{\gamma}_\alpha^{\mathrm{c}}\right| \tag{33}$$

The slip system exhibiting the highest activity is selected to determine the current crack propagation direction:

$$\alpha^* = \arg\max_\alpha A_\alpha \tag{34}$$

The crack plane is assumed to be aligned with the currently most active slip direction, with the crack normal oriented perpendicular to the slip direction.

Once crack initiation occurs, the subsequent crack propagation is assumed to be controlled by the intragranular fracture energy, which is defined as follows:

$$G_{\mathrm{c}}^{\mathrm{in}} = \int T_{\mathrm{in}} \mathrm{d}\delta_{\mathrm{in}} \tag{35}$$

where $T_{\mathrm{in}}$ and $\delta_{\mathrm{in}}$ denote the equivalent traction and separation across the intragranular crack surface (see Fig. 4b), respectively. In the present work, an additional intragranular damage variable is not introduced to degrade the bulk stress in the crystal plasticity formulation. This treatment avoids potential double counting of energy dissipation arising from crystal plasticity slip, XFEM-based crack propagation, and GB cohesive degradation.

## 3.4. Energy decomposition and thermodynamic admissibility

From an energy perspective, this section examines the partition of energy dissipation, the evolution of internal variables, and the thermodynamic admissibility of the proposed framework. The energy analysis is intended to demonstrate the role of each dissipative mechanism and to ensure that fracture energy is accounted for only once during crystal plasticity deformation, GB decohesion, and XFEM-based intragranular crack propagation.

The total Helmholtz free energy consists of the bulk crystal plasticity contribution, the GB cohesive contribution, and the intragranular cohesive contribution associated with XFEM crack surfaces.

$$\Psi = \int_{\Omega} \psi_{\mathrm{cp}} \mathrm{d}V + \int_{\Gamma_{\mathrm{GB}}} \psi_{\mathrm{GB}} \mathrm{d}A + \int_{\Gamma_{\mathrm{in}}} \psi_{\mathrm{XFEM}}^{\mathrm{in}} \mathrm{d}A \tag{36}$$

The crystal plasticity free energy density $\psi_{\mathrm{cp}}$ accounts for the elastic deformation of the grains and the evolution of the internal hardening variables. The GB free energy density $\psi_{\mathrm{GB}}$ depends on the recoverable interfacial separation and the GB damage variable $D_{\mathrm{GB}}$, whereas the intragranular XFEM free energy density $\psi_{\mathrm{XFEM}}^{\mathrm{in}}$ is associated with the separation of the intragranular crack surfaces. The GB free energy can be expressed as:

$$\psi_{\mathrm{GB}} = \frac{1}{2} g\left(D_{\mathrm{GB}}\right) k_{\mathrm{n}} u_{\mathrm{n}}^2 + \frac{1}{2} g\left(D_{\mathrm{GB}}\right) k_{\mathrm{t}} u_{\mathrm{t}}^2 \tag{37}$$

The total dissipation can be formally decomposed as:

$$\mathcal{D}_{\mathrm{tot}} = \mathcal{D}_{\mathrm{cp}} + \mathcal{D}_{\mathrm{GB}} + \mathcal{D}_{\mathrm{XFEM}}^{\mathrm{in}} \geq 0 \tag{38}$$

The crystal plasticity dissipation, $\mathcal{D}_{\mathrm{cp}}$, is associated with crystal slip and hardening. The GB dissipation, $\mathcal{D}_{\mathrm{GB}}$, arises from irreversible GB opening induced by cavity nucleation, growth and coalescence, together with GB tangential sliding and degradation of the remaining ligaments. Since GB degradation is formulated in terms of a damage variable, the thermodynamic driving force conjugate to the damage evolution is given by:

$$Y_{\mathrm{D}} = -\frac{\partial \psi_{\mathrm{GB}}}{\partial D_{\mathrm{GB}}} \tag{39}$$

Accordingly, the GB dissipation can be expressed as:

$$\mathcal{D}_{\mathrm{GB}} = Y_{\mathrm{D}} \dot{D}_{\mathrm{GB}} \geq 0 \tag{40}$$

Since the degradation function satisfies $g'\left(D_{\mathrm{GB}}\right) \leq 0$ for $0 < D_{\mathrm{GB}} \leq 1$, and the damage evolution is irreversible ($\dot{D}_{\mathrm{GB}} \geq 0$), the GB dissipation remains non-negative, thereby satisfying the Clausius-Duhem inequality.

The intragranular XFEM dissipation, $\mathcal{D}_{\mathrm{XFEM}}^{\mathrm{in}}$, is associated with cohesive separation across the trans-granular crack surfaces. Because the trans-granular XFEM model is likewise formulated using a cohesive traction-separation law, its dissipation can be derived in the same manner as that of the GB cohesive model. Nevertheless, the two models represent different fracture processes, with GB CZM describing inter-granular decohesion and XFEM describing trans-granular crack evolution. The distinction involves both the crack initiation condition and the subsequent damage evolution mechanism. The dissipation rate associated with trans-granular cracking can be related to the trans-granular fracture energy as follows:

$$G_{\mathrm{XFEM}}\left(\delta_{\mathrm{in}}\right) = \int_0^{\delta_{\mathrm{in}}} \mathcal{D}_{\mathrm{XFEM}}^{\mathrm{in}}\left(\delta\right) \mathrm{d}\delta \tag{41}$$

and

$$G_{\mathrm{c}}^{\mathrm{in}} = G_{\mathrm{XFEM}}\left(\delta_{\mathrm{c}}\right) \tag{42}$$

in which $\delta_{\mathrm{c}}$ represents the critical separation displacement corresponding to the complete opening of the trans-granular crack.

The present framework adopts the following energy partition strategy. The crystal plasticity formulation is employed to describe grain deformation, including crystal slip and hardening evolution, without introducing an additional bulk damage variable to degrade the CP stress response. GB degradation is described through the cohesive damage variable $D_{\mathrm{GB}}$, which accounts for irreversible inter-granular decohesion. Meanwhile, the fracture energy $G_{\mathrm{c}}^{\mathrm{in}}$ in the XFEM formulation is exclusively assigned to trans-granular crack propagation within grains and is not involved in GB degradation.

It should be noted that the present model does not establish a unified diffusion-based framework in which vacancy transport simultaneously governs bulk creep deformation and GB damage evolution. Instead, these two processes are described by separate constitutive formulations, and their interaction is mediated through the mechanical fields of the coupled model. Consequently, the dissipative contributions associated with crystal plasticity, GB degradation, and trans-granular crack propagation remain clearly partitioned. Each dissipation mechanism corresponds to a distinct physical process, thereby avoiding double counting of fracture-related energy dissipation.

## 3.5. Computational scheme and calibration

The formulations presented in Sections 3.1~3.4 are implemented within a commercial finite element software ABAQUS. The crystal plasticity constitutive formulation is implemented through a user-defined material subroutine (UMAT), the GB cohesive model is implemented through a user-defined element subroutine (UEL), and the trans-granular crack initiation criterion is incorporated through the user-defined damage initiation subroutine (UDMGINI). This finite element framework enables the coupled simulation of creep deformation, inter-granular failure, and trans-granular failure within a unified computational framework.

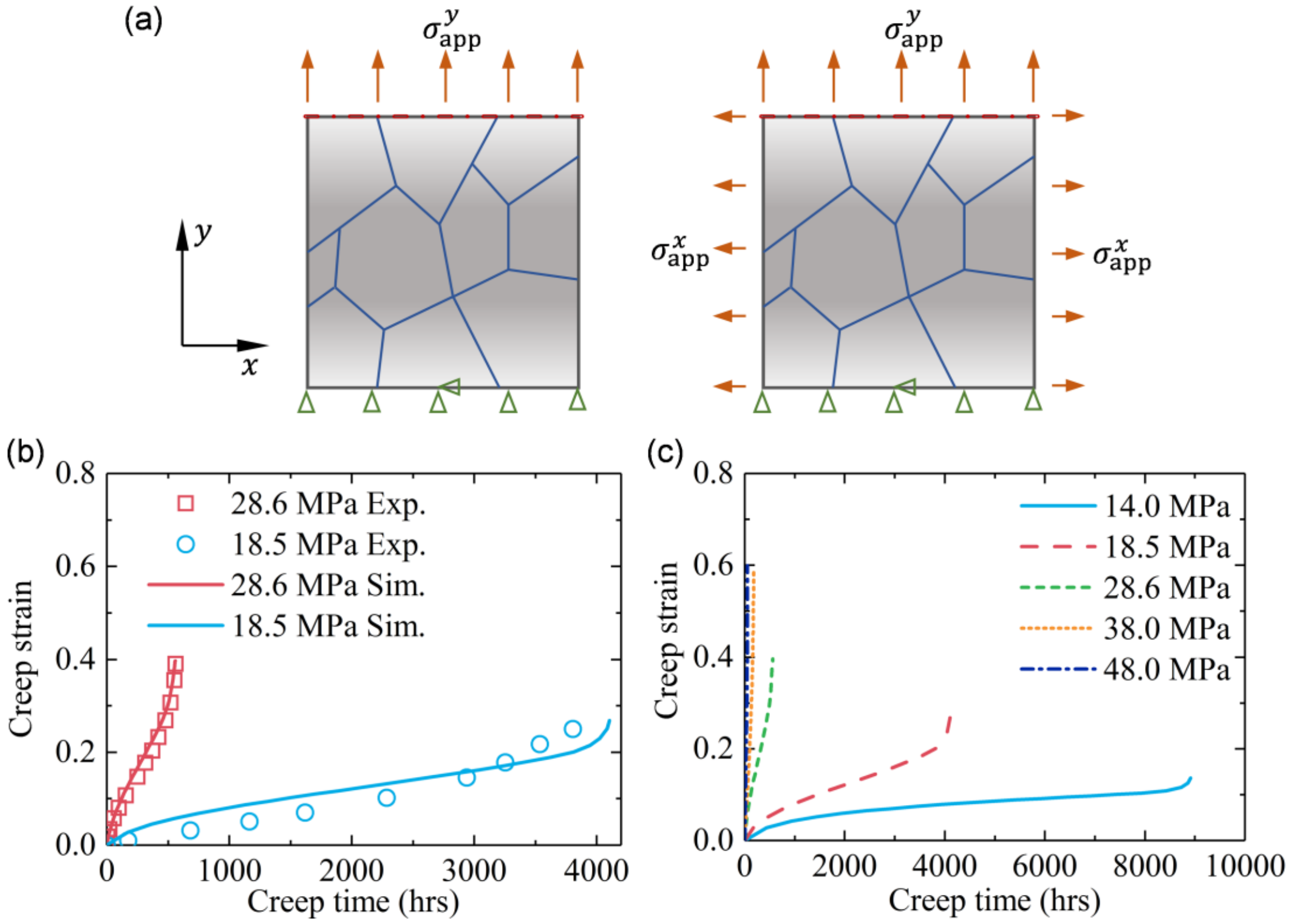

Fig. 5 (a) Schematic illustration of the loading conditions applied to the polycrystalline RVE under uniaxial and multiaxial loading; (b) Comparison of the equivalent macroscopic creep strain curves predicted by the RVE model with the experimental ones; (c) Equivalent macroscopic creep strain curves under different applied stress levels.

As shown in Fig. 5a, a two-dimensional (2D) polycrystalline plane model is employed in the present study. The grain size distribution follows the statistical characterization reported by Zhang and Oskay (2016), which is obtained from electron backscatter diffraction (EBSD) measurements of Inconel 617. Random crystallographic orientations are then assigned to the grains to represent the polycrystalline texture.

To capture damage-induced localization and failure within a finite polycrystalline domain, a mixed boundary loading condition is adopted, where partial displacement constraints are imposed together with constant traction loading. As shown in Fig. 5a, the bottom boundary of the finite polycrystalline domain is constrained in the $y$-direction, with additional constraints imposed to eliminate rigid-body motion. A constant tensile load $\sigma^y_{app}$ is applied to the top boundary. In addition, a constraint equation is imposed to enforce identical $y$-direction displacements for all finite element nodes along the top boundary, such that the top boundary remains planar throughout the tensile creep process. For the multiaxial loading case, an additional tensile stress $\sigma_{\mathrm{app}}^{x}$ is applied. Failure of the finite polycrystalline domain is identified from the evolution of the macroscopic creep curve. As failure is approached, the creep strain rate exhibits a sudden acceleration, accompanied by a sharp increase in displacement within a single time increment. Beyond this point, the structural response becomes unstable and the equilibrium solution can no longer be robustly obtained. Therefore, the converged solution at the preceding time increment is taken as the final state immediately prior to failure. A finite polycrystalline domain with a side length of 600 μm is employed for material parameter calibration. As shown in Fig. 5b, the crystal plasticity, GB cohesive, and trans-granular fracture parameters are calibrated against the experimental creep data of Inconel 617 at 950 °C reported by Natesan et al. (2012). The calibrated model parameters are summarized in Table 1~3. Using the calibrated model parameters, creep simulations are performed under different applied stress levels to obtain the corresponding macroscopic creep strain curves (see Fig. 5c).

Table 1 Crystal plasticity constitutive parameters

| Parameter | Value | Parameter | Value |
|---|---|---|---|
| $C_{11}$ (GPa) | 170.64 | $p_g$ | 0.181 |
| $C_{12}$ (GPa) | 108.39 | $q_g$ | 1.633 |
| $C_{44}$ (GPa) | 77.82 | $h_{\mathrm{S}}$ (MPa) | 397.73 |
| $\dot{\gamma}_{0\mathrm{c}}$ ($\mathrm{s}^{-1}$) | $1.299\times10^{-3}$ | $d_{\mathrm{D}}$ (MPa) | 5073.62 |
| $F_0$ (J) | $5.148\times10^{-19}$ | $S_0^{\alpha}$ (MPa) | 143.41 |
| $\mu_0$ (GPa) | 265.33 | $\bar{S}^{\alpha}$ (MPa) | 18.03 |
| $\mu$ (GPa) | 77.82 | $h_{\mathrm{B}}$ (MPa) | 104.31 |
| $\mu_0'$ | 31.13 | $h_2$ (MPa) | 0.015 |
| $\hat{\tau}_0$ (MPa) | 268.2 | $\dot{\gamma}_{\mathrm{th}}$ ($\mathrm{s}^{-1}$) | $1.0\times10^{-6}$ |
| $f_{\mathrm{co}}$ | 0.36 | $\psi_{\mathrm{c}}$ (rad) | $\pi/4$ |
| $p_{\mathrm{c}}$ | 3.0 | $\hat{\tau}_{0\mathrm{c}}$ (Pa) | 7350 |
| $h_{\mathrm{c}}$ | 32 | $B_{\infty}$ (MPa) | 3.0 |

Table 2 Material parameters for the GB cohesive zone model

| Parameter | Value | Parameter | Value |
|---|---|---|---|
| $\psi_{\mathrm{cav}}$ (°) | 70.0 | $a_{\mathrm{i}}$ (μm) | 0.5 |
| $k_{\mathrm{n}}$ (GPa) | $1.0\times10^{5}$ | $N_{\mathrm{i}}$ ($\mu\mathrm{m}^{-2}$) | $4.0\times10^{-4}$ |
| $k_t$ (GPa) | $5.0\times10^{4}$ | $\alpha$ | 0.5 |
| $k_c$ (GPa) | $1.0\times10^{6}$ | $A_D\left(s^{-1}\right)$ | $2.04\times10^{-3}$ |
| $\delta_c\,(\mu m)$ | 450.0 | $r_D$ | 0.5 |
| $T_N^c\,(MPa)$ | 300.0 | $m_D$ | 1.0 |
| $T_N\,(MPa)$ | 800.0 | $A_C\left(s^{-1}\right)$ | $8.38\times10^{-13}$ |
| $T_N^0\,(MPa)$ | 800.0 | $r_C$ | 1.2 |
| $N_{max}\left(\mu m^{-2}\right)$ | $1.2\times10^{-2}$ | $m_C$ | 5.24 |
| $A_N\left(s^{-1}\right)$ | $2.0\times10^{-12}$ | $T_c\,(MPa)$ | 300.0 |
| $p_N$ | 1.0 | $T_0\,(MPa)$ | 800.0 |
| $q_N$ | 1.7 | $\chi_i$ | 0.01 |
| $b_i\,(\mu m)$ | 50.0 | $\chi_c$ | 0.90 |
| $\dot{V}_i\left(\mu m^3\right)$ | 0.32 | | |

Table 3 Material parameters for the trans-granular fracture model

| Parameter | Value | Parameter | Value |
|---|---|---|---|
| $\Gamma_c$ | 1.67 | $G_c^{in}\,(Pa\cdot\mu m)$ | 300 |

# 4. Stress-dependent trade-off in creep fracture transition

The unified crystal plasticity-fracture framework, which simultaneously incorporates crystal plasticity, GB damage, and trans-granular fracture, enables a systematic investigation of creep cracking over a wide range of applied stress levels. The framework provides predictions of the macroscopic creep strain response and creep life, and is further applied to examine the creep cracking mechanisms, macroscopic ductility, and creep life under multiaxial stress states.

## 4.1. Stress-sensitive evolution of fracture mechanisms

Based on the proposed crystal plasticity-fracture framework, the creep cracking behaviour of Inconel 617 is simulated under different applied stress levels. Fig. 6a presents the evolution of creep cracking under a constant applied stress of 18.5 MPa. At relatively low applied stress levels, the microstructure remains largely intact over an extended period. After approximately 3500 h of creep exposure, multiple inter-granular cracks begin to nucleate along the GBs. As shown in Fig. 6b, inter-granular cracks typically nucleate at wedge regions and subsequently propagate along the adjacent GBs in both directions. As creep proceeds, inter-granular crack initiation becomes increasingly distributed throughout the microstructure, while localized crack coalescence gradually develops in several regions. As shown in Fig. 6c~d, the coalescence of neighbouring short cracks eventually leads to localized inter-granular damage and the formation of a dominant crack. This localization process occurs during the final stage of the creep life. For the case of 18.5 MPa, it develops mainly between approximately 3800 and 4100 h, corresponding to the last 7% of the total creep life. During this stage, the dominant crack propagates rapidly and the macroscopic strain increases markedly, corresponding to the accelerated deformation characteristic of tertiary creep. It is also worth noting that no pronounced macroscopic necking is observed prior to failure. Instead, deformation remains highly localized around the crack path, indicating that the fracture process is dominated by localized cracking rather than extensive plastic instability. This macroscopic fracture characteristic is consistent with the post-fracture morphology observed experimentally in the cylindrical specimens (Wang et al., 2022). The simulated inter-granular fracture morphology is in good qualitative agreement with the experimental fracture surface shown in Fig. 2b, exhibiting the characteristic faceted appearance associated with inter-granular fracture. Meanwhile, as shown in Fig. 6e, locally elevated instantaneous stresses in the failure region can induce a small number of trans-granular cracks. These cracks are mainly associated with local damage accompanying inter-granular crack propagation and therefore play only a secondary role in the overall failure process under low-stress creep conditions.

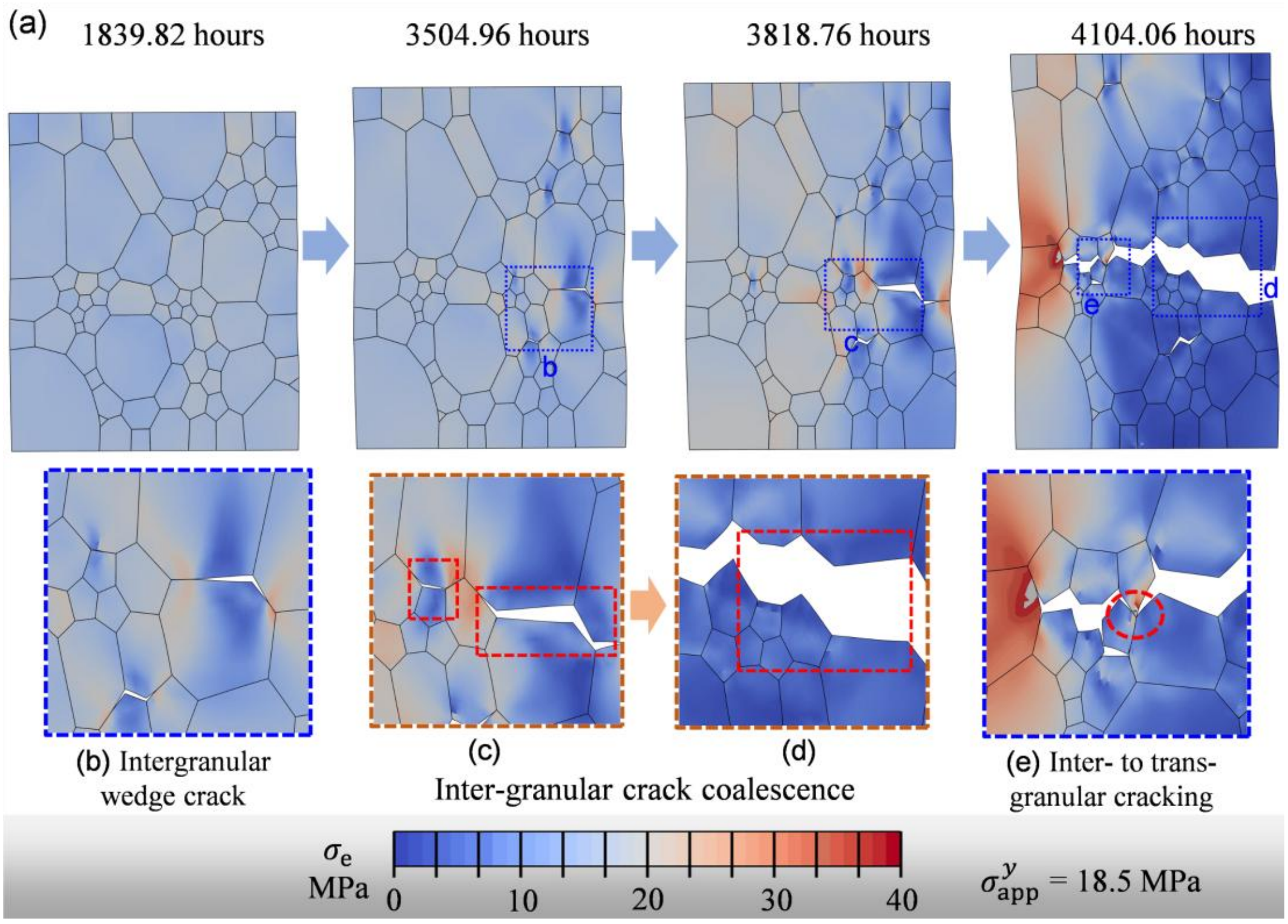

Fig. 6 Creep cracking evolution under an applied stress of 18.5 MPa: (a) overall evolution of creep cracking; (b)~(e) enlarged views of the local crack morphology at different stages

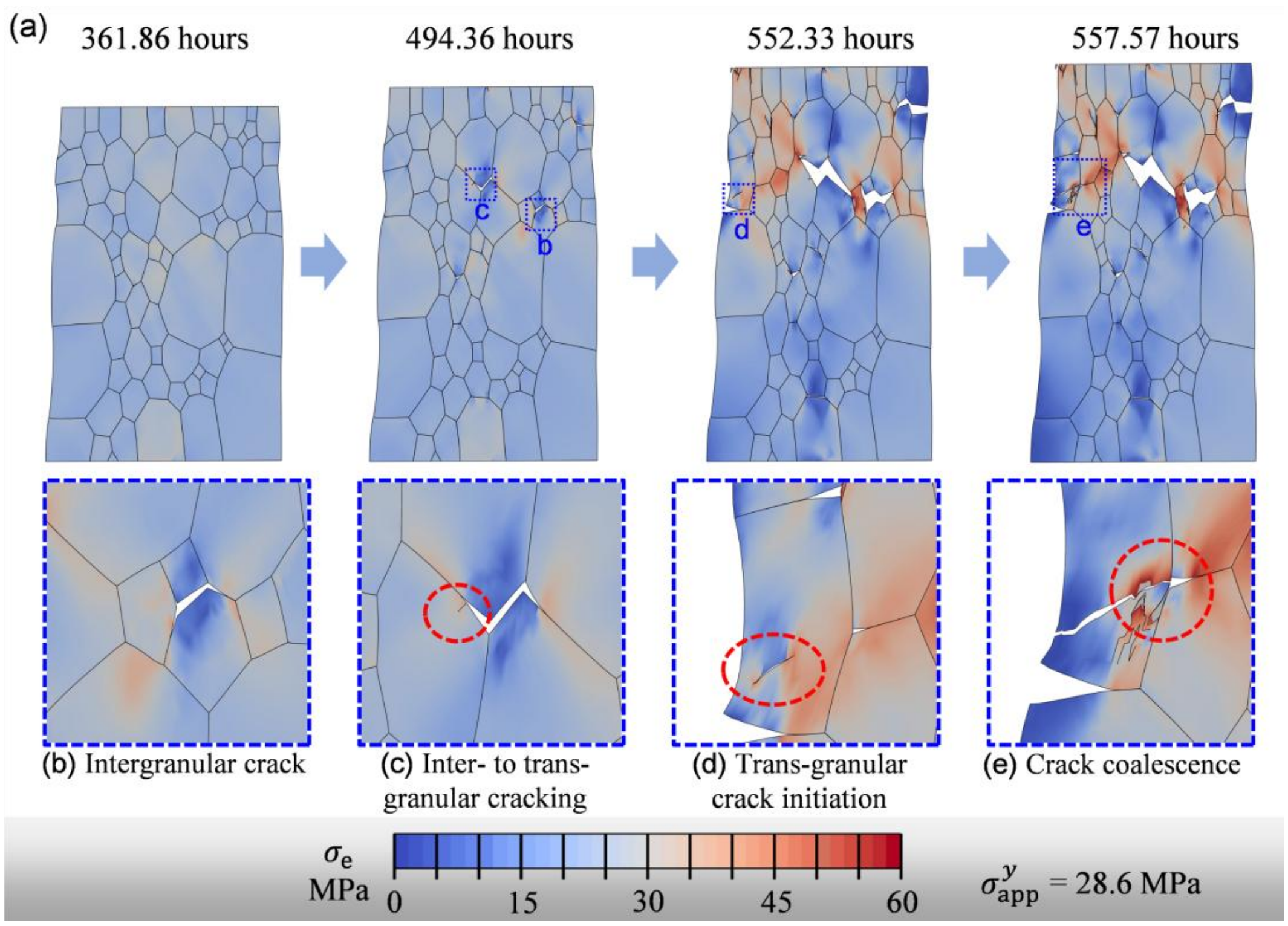

Fig. 7 Creep cracking evolution under an applied stress of 28.6 MPa: (a) overall evolution of creep cracking; (b)~(e) enlarged views of the local crack morphology at different stages

Fig. 7a presents the evolution of creep damage and cracking under a constant applied stress of 28.6 MPa. Under this stress level, failure is also initiated by inter-granular cracking. As shown in Fig. 7b, typical inter-granular cracks develop along the GBs. Besides, Fig. 7c shows that localized trans-granular cracks are induced in the vicinity of the propagating inter-granular cracks, indicating that cracks can propagate from the GBs into the grain interiors. In addition to the trans-granular cracks that develop in association with propagating inter-granular cracks, another type of trans-granular cracking is observed under this stress level. As shown in Fig. 7d, these cracks nucleate directly within the grain interiors and subsequently propagate without being initiated from the GBs, representing another important mode of creep damage. As shown in Fig. 7e, trans-granular cracks nucleate and propagate along the directions of the currently most active slip systems within the grain interiors. Coalescence between trans-granular and inter-granular cracks is also observed. At an applied stress of 28.6 MPa, pronounced macroscopic deformation develops prior to failure, while no significant necking is observed. Compared with the low-stress case, the material exhibits a more ductile deformation response before fracture, yet the absence of noticeable necking indicates a quasi-brittle fracture characteristic at the macroscopic scale.

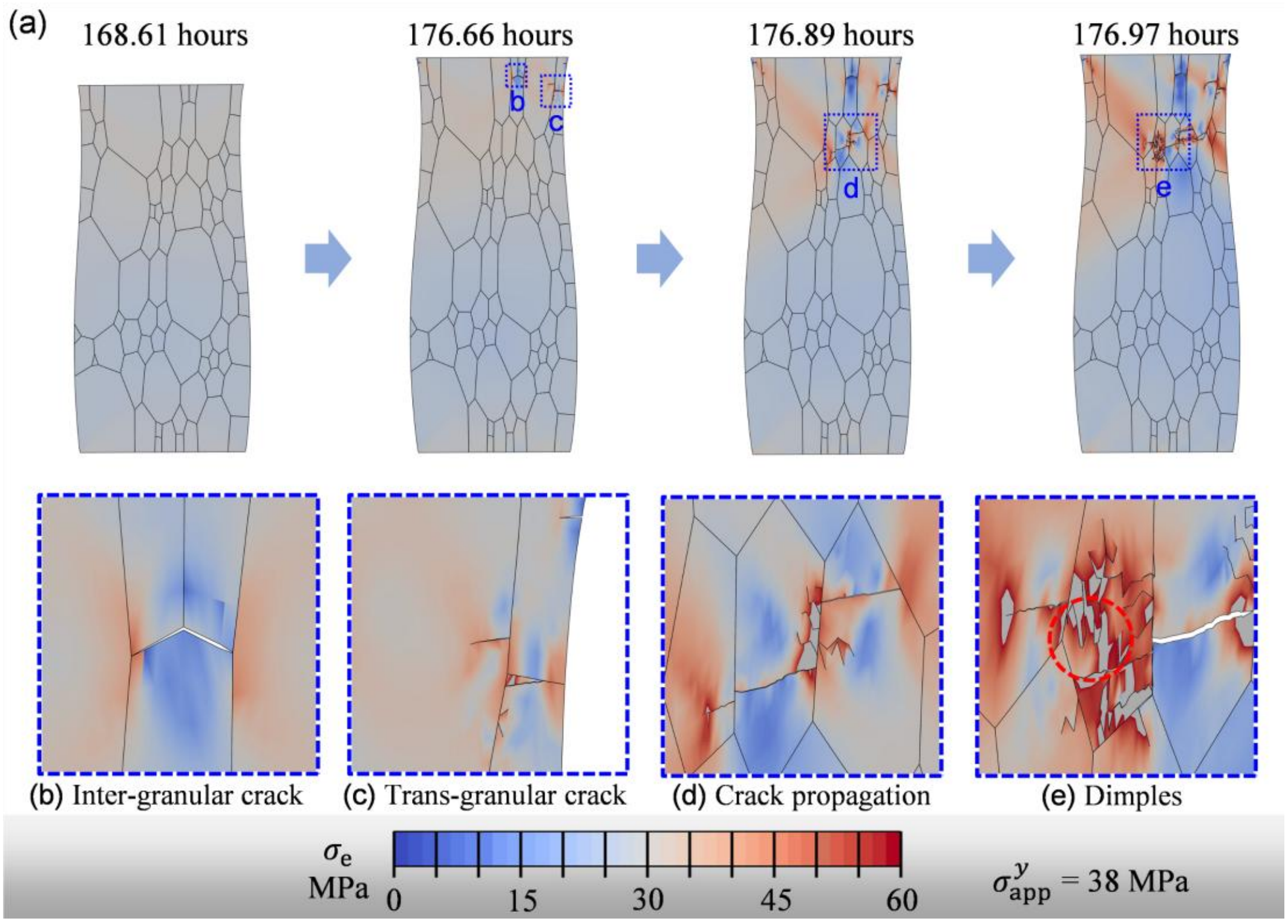

Fig. 8 Creep cracking evolution under an applied stress of 38 MPa: (a) overall evolution of creep cracking; (b)~(e) enlarged views of the local crack morphology at different stages

Fig. 8a presents the evolution of creep cracking under a constant applied stress of 38 MPa. Compared with the lower stress levels, a distinctly different cracking behaviour is observed. The results indicate that substantial plastic deformation develops prior to the onset of observable cracking. As shown in Fig. 8b~c, inter-granular and trans-granular cracks develop concurrently during creep. Trans-granular cracks preferentially nucleate near the free surface and in the vicinity of highly stressed GBs, and subsequently propagate into the grain interiors. As creep proceeds, trans-granular cracks rapidly propagate under the 38 MPa loading condition, forming pronounced tortuous crack paths along the active slip systems, as shown in Fig. 8d. By comparison, the GB cracks exhibit only limited propagation during this stage. In addition, regions with a high crack density exhibit localized clusters of slip-induced microcracks, giving rise to a dimple-like damage morphology, as shown in Fig. 8e. This damage morphology is in good qualitative agreement with the experimental fracture surface shown in Fig. 2a, where a high density of fine intragranular dimples is observed. From the macroscopic perspective, pronounced creep deformation and necking develop under the relatively high constant applied stress, indicating a clear transition toward a ductile fracture response. Under these creep conditions, substantial creep deformation and necking precede the final structural instability and fracture. The observation is in good qualitative agreement with the macroscopic fracture morphology reported by Wang et al. (2022).

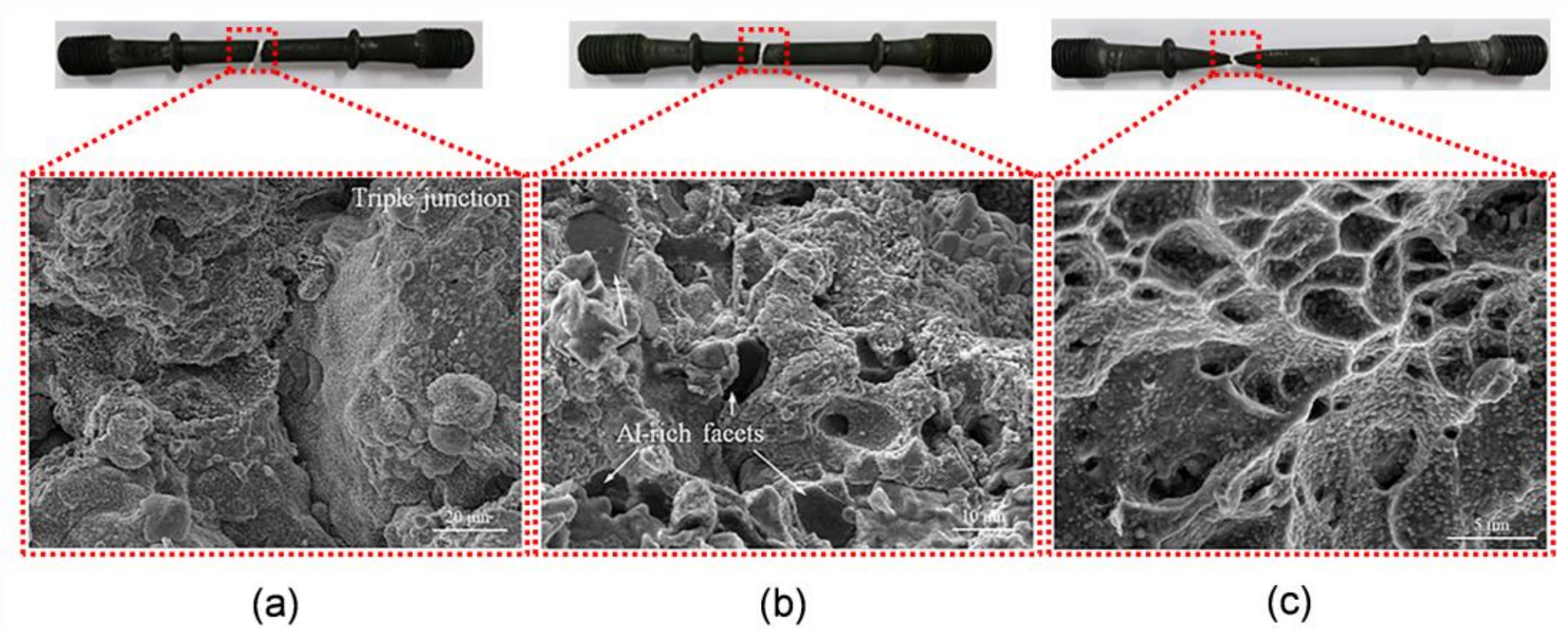

Fig. 9 Experimental observations of creep fracture morphologies of Inconel 617 under uniaxial loading: (a) inter-granular cracking-dominated failure at 1000 °C and 16 MPa; (b) mixed-mode failure at 1000 °C and 30 MPa; (c) trans-granular cracking-dominated failure at 800 °C and 80 MPa.

As shown in Fig. 9, the fracture morphologies of Inconel 617 under uniaxial constant-stress creep conditions illustrate the transition of microscopic creep cracking mechanisms from inter-granular cracking-dominated failure to mixed-mode and trans-granular cracking-dominated failure. As shown in Fig. 9a, under low-stress creep conditions, the fracture surface exhibits a characteristic rock-like morphology, indicating that the failure is dominated by inter-granular cracking induced by GB damage. In Fig. 9b, a mixed fracture morphology involving both inter-granular and trans-granular cracking is observed, where rock-like features associated with GB fracture coexist with intragranular dimples. In contrast, Fig. 9c presents a creep fracture morphology dominated entirely by trans-granular cracking, characterized by dense dimples with sizes smaller than the average grain size. The transition of microscopic creep cracking mechanisms is reflected macroscopically by the evolution of failure behaviour from quasi-brittle to ductile rupture, as illustrated in Fig. 9. The experimentally observed transition in fracture morphology agrees well with the simulation results obtained from the proposed framework (Figs. 6~8), demonstrating its capability to capture the stress-dependent evolution of creep fracture behaviour.

It should be noted that this transition is governed not only by the applied stress level but also strongly influenced by temperature, as demonstrated by the creep fracture mechanism maps established by Ashby and co-workers (Ashby et al., 1979). In the present study, experimental observations under different stress-temperature conditions are used to validate the reliability of the proposed framework. Although the current simulations are conducted at a single temperature condition, the proposed framework provides a general methodology for investigating creep fracture transitions under different temperature regimes, provided that the temperature-dependent constitutive parameters are appropriately calibrated.

## 4.2. Micromechanical origin of the transition from inter- to trans-granular cracking

To quantify the influence of creep stress level on the creep cracking mechanism, a series of simulations are performed using randomly generated polycrystalline microstructures. The relative proportions of trans-granular and inter-granular cracks at failure are then statistically evaluated to reduce the influence of individual microstructural realizations and provide a statistically robust characterization of the stress-dependent cracking behaviour. Constant-stress creep simulations are conducted at applied stress levels of 14, 18.5, 28.6, 38, and 48 MPa. For each stress level, 10 randomly generated polycrystalline domains with a side length of 300 μm are considered to account for microstructural variability and to support the statistical evaluation of inter-granular and trans-granular cracking at failure.

The crack morphologies at failure for the ten random polycrystalline domains under the five applied stress levels are presented in Fig. 10 , 12, 14, 16, and 18, respectively. With increasing applied stress, the failure behaviour progressively changes from predominantly brittle-like to ductile fracture. Correspondingly, the local damage mode shifts from inter-granular cracking to trans-granular ductile damage. This transition trend is consistently reproduced across the 10 random polycrystalline domains at each stress level, demonstrating its statistical robustness rather than being an isolated microstructural response.

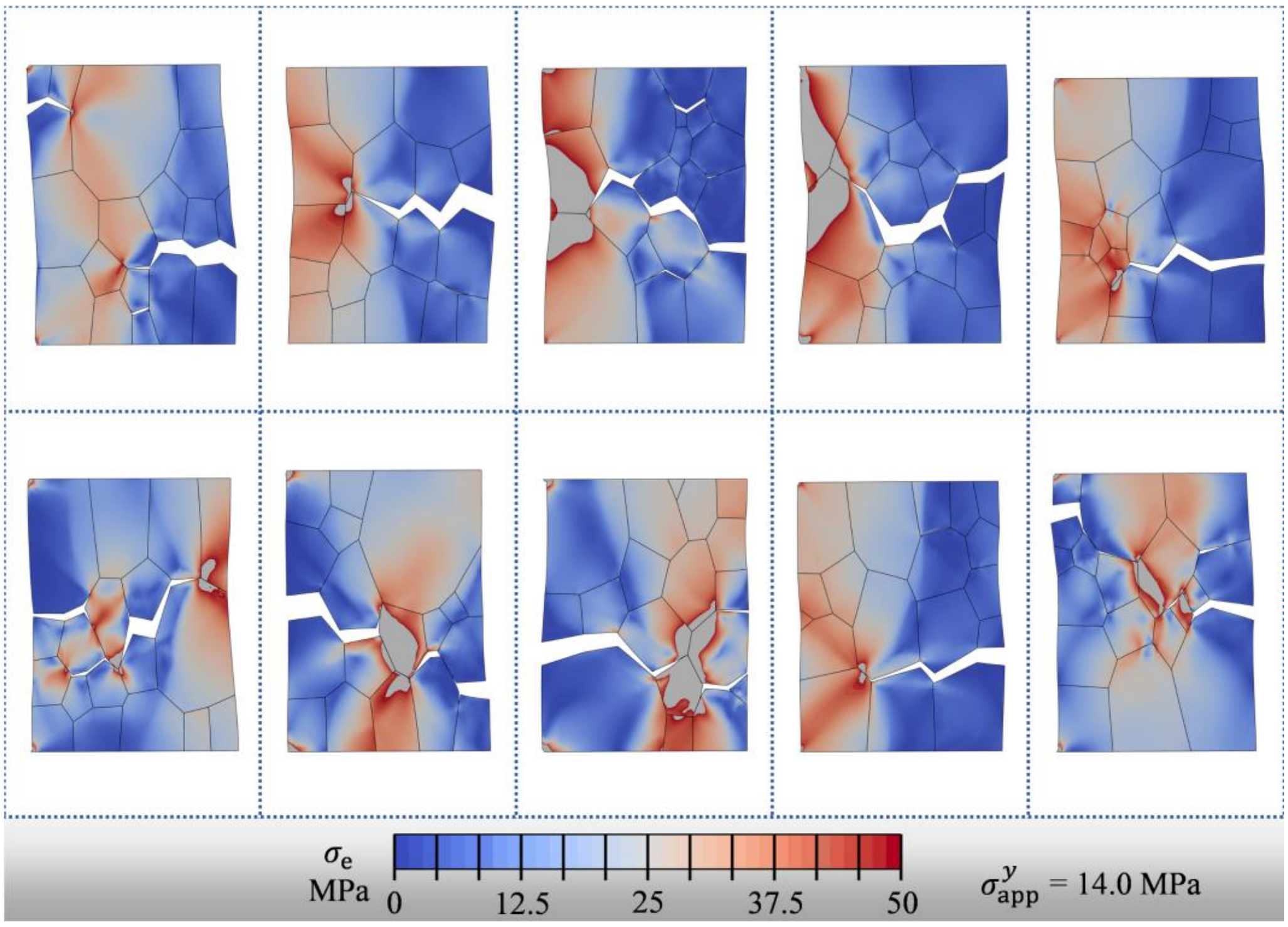

Fig. 10 Creep failure morphology of a random polycrystalline domains with a side length of 300 μm under uniaxial constant-stress creep loading at 14 MPa.

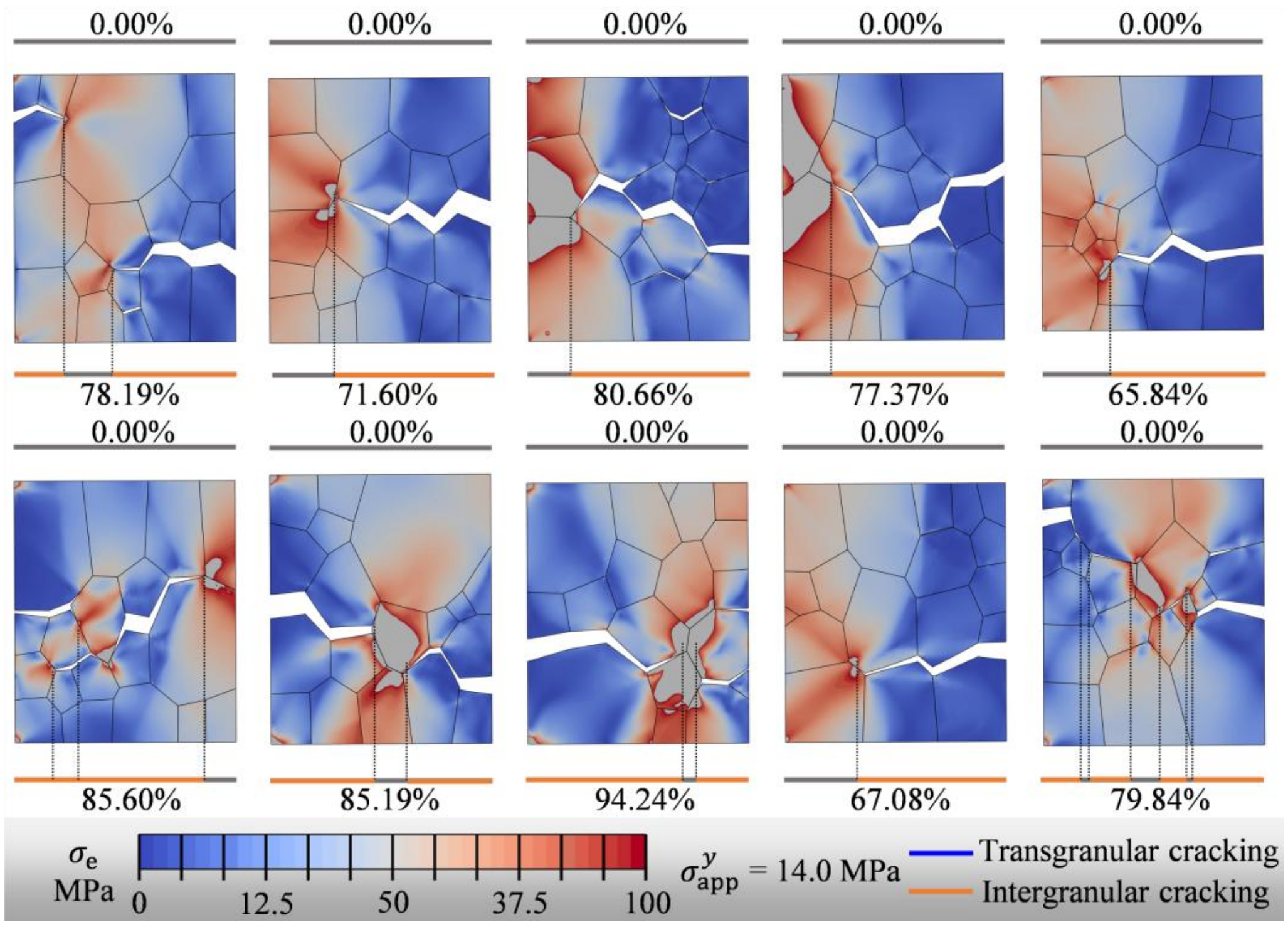

Fig. 11 Normalized lengths of inter-granular and trans-granular cracks with respect to the initial

width of the polycrystalline domain, for a random polycrystalline domain with a side length of 300 μm under uniaxial constant-stress creep loading at 14 MPa.

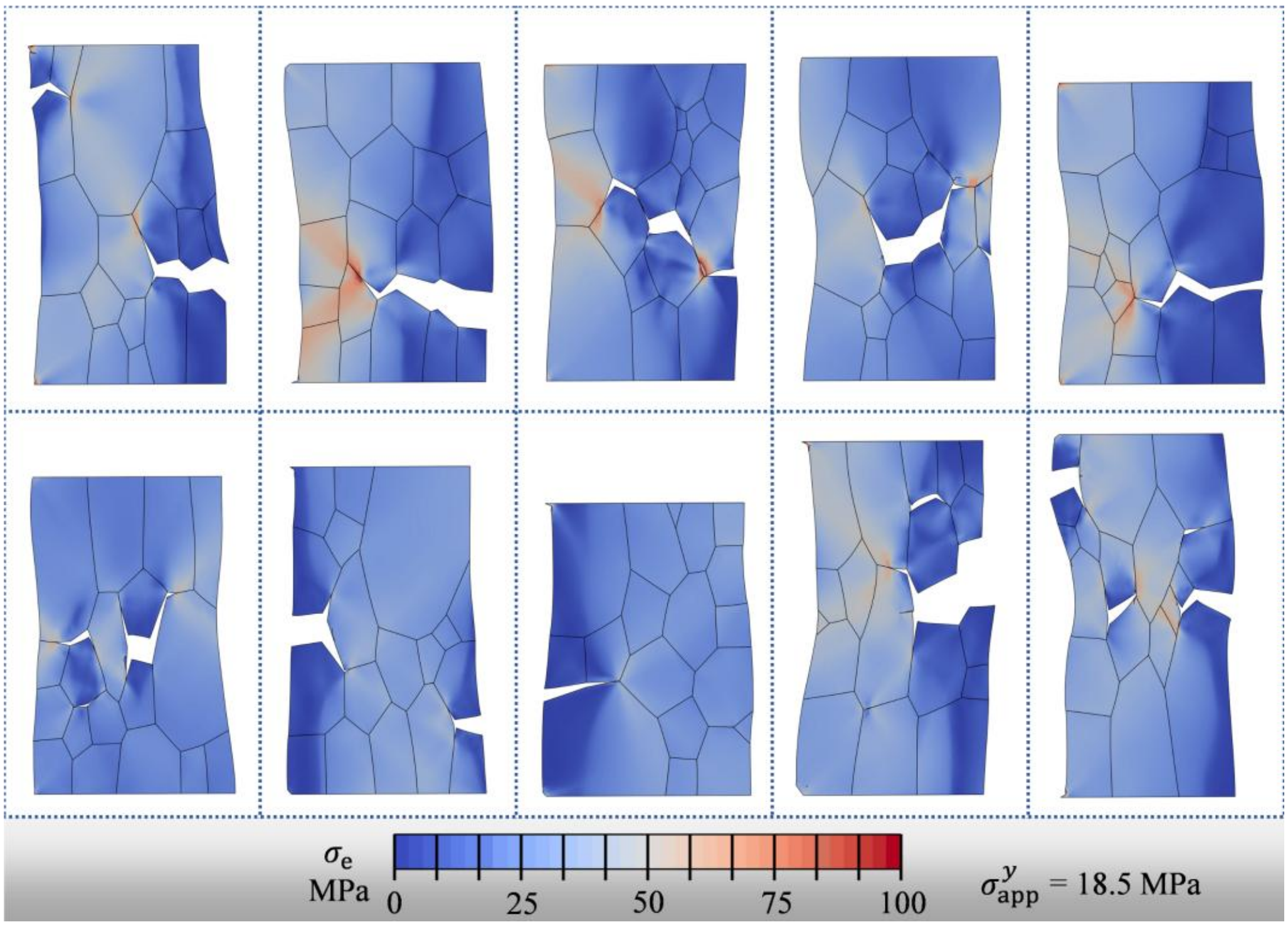

Fig. 12 Creep failure morphology of a random polycrystalline domains with a side length of 300 μm under uniaxial constant-stress creep loading at 18.5 MPa.

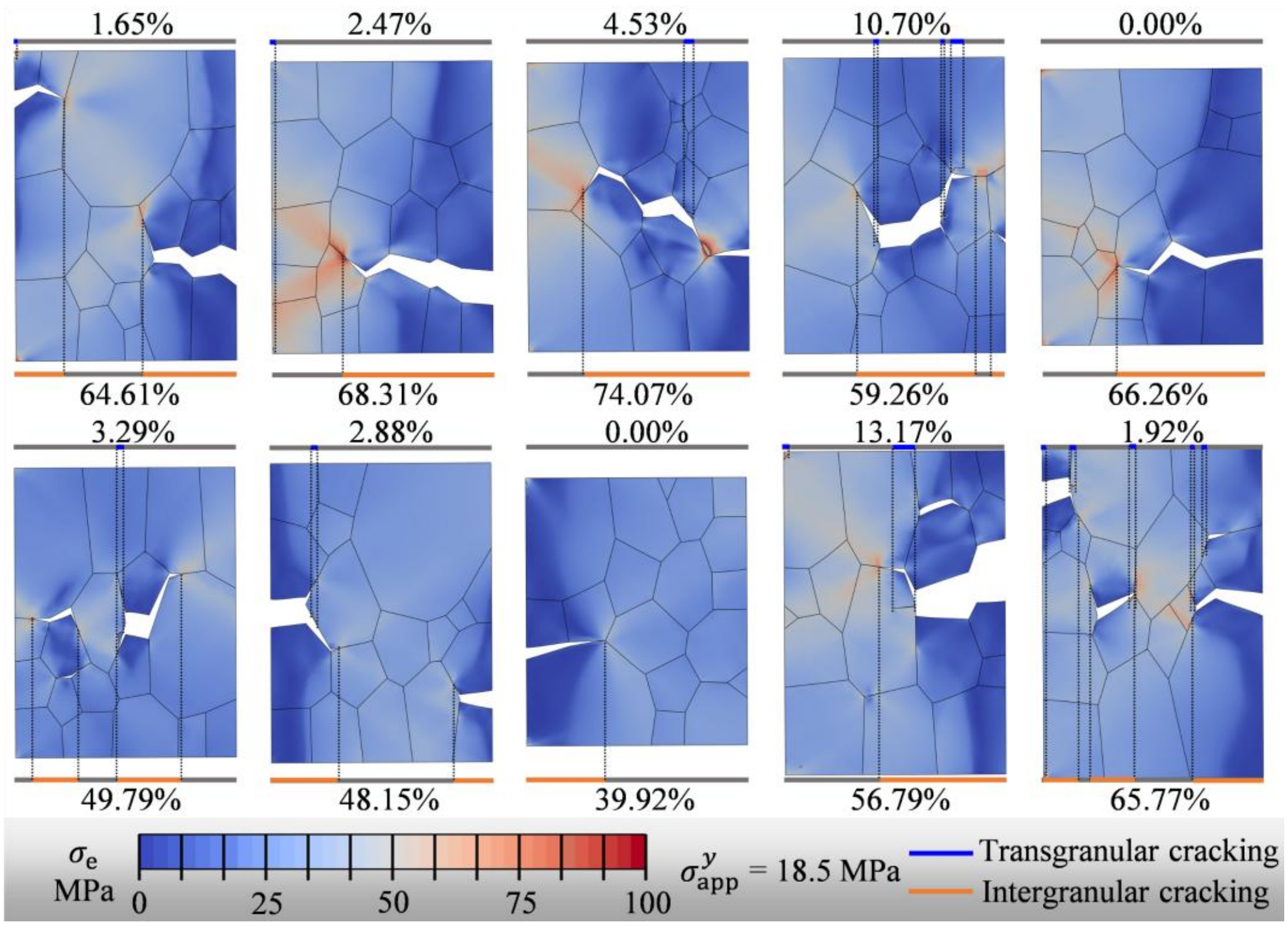

Fig. 13 Normalized lengths of inter-granular and trans-granular cracks with respect to the initial width of the polycrystalline domain, for a random polycrystalline domain with a side length of 300 μm

under uniaxial constant-stress creep loading at 18.5 MPa.

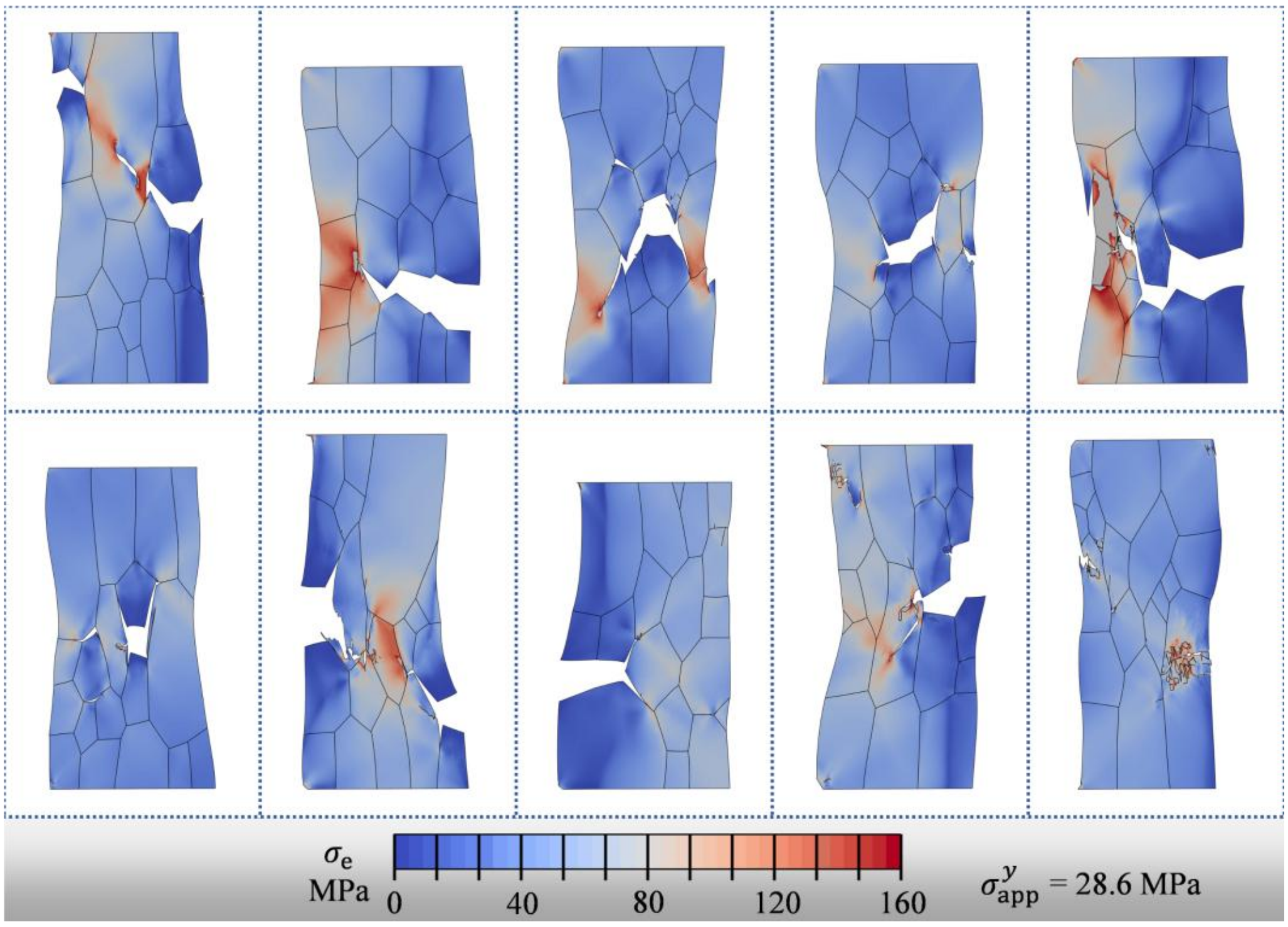

Fig. 14  Creep failure morphology of a random polycrystalline domains with a side length of 300 μm under uniaxial constant-stress creep loading at 28.6 MPa.

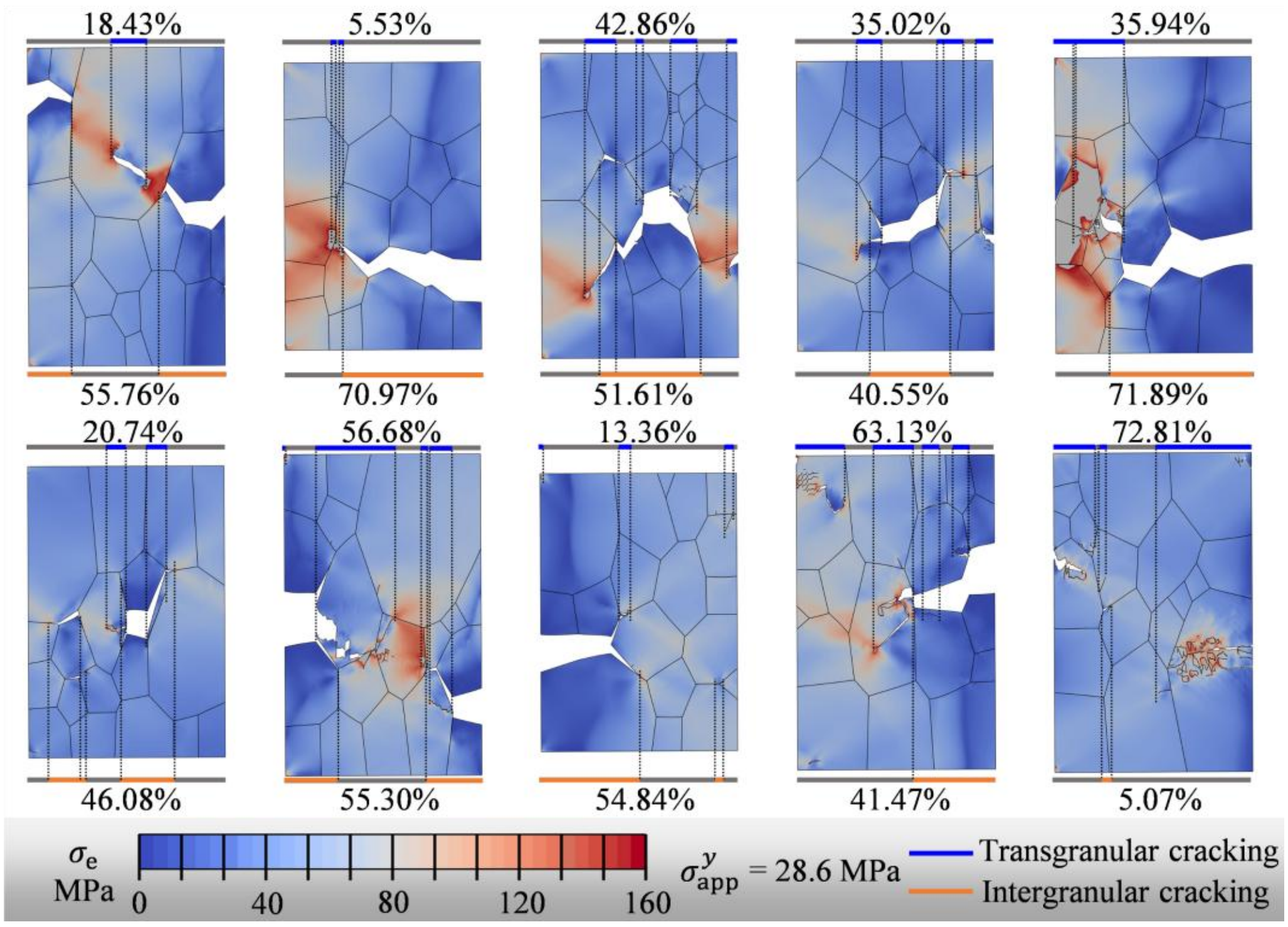

Fig. 15  Normalized lengths of inter-granular and trans-granular cracks with respect to the initial width of the polycrystalline domain, for a random polycrystalline domain with a side length of 300 μm under uniaxial constant-stress creep loading at 28.6 MPa.

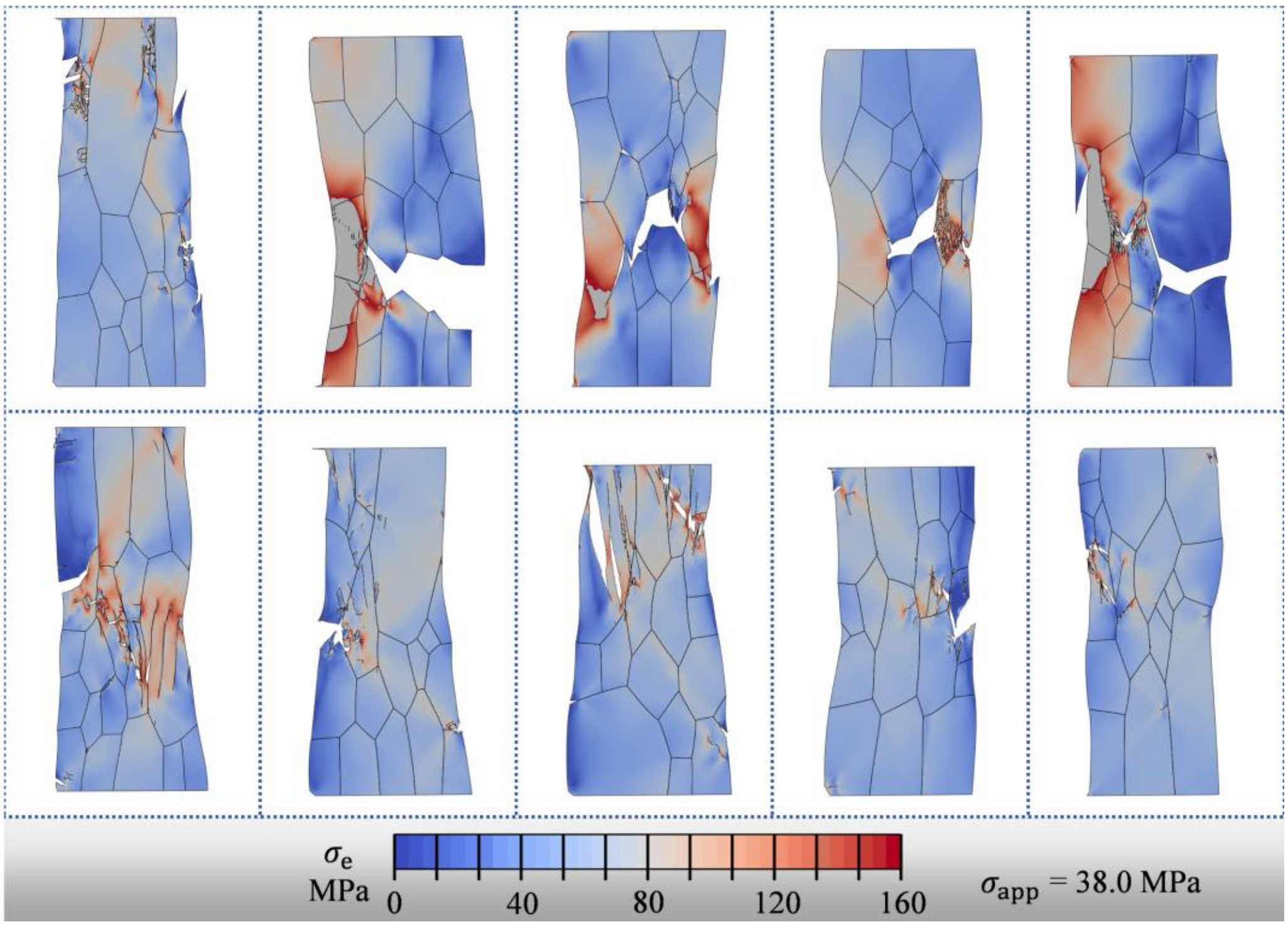

Fig. 16 Creep failure morphology of a random polycrystalline domains with a side length of 300 μm under uniaxial constant-stress creep loading at 38 MPa.

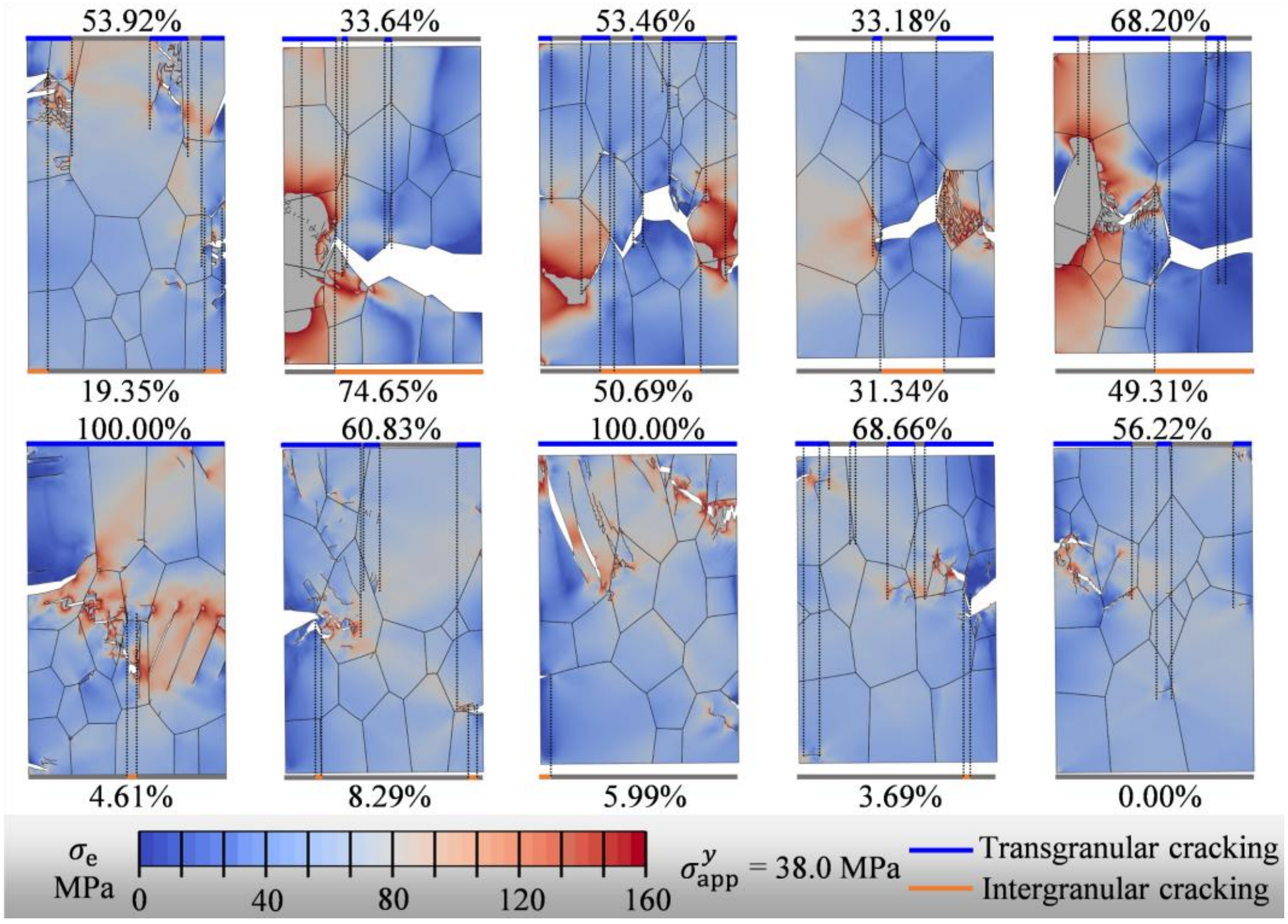

Fig. 17 Normalized lengths of inter-granular and trans-granular cracks with respect to the initial width of the polycrystalline domain, for a random polycrystalline domain with a side length of 300 μm under uniaxial constant-stress creep loading at 38 MPa.

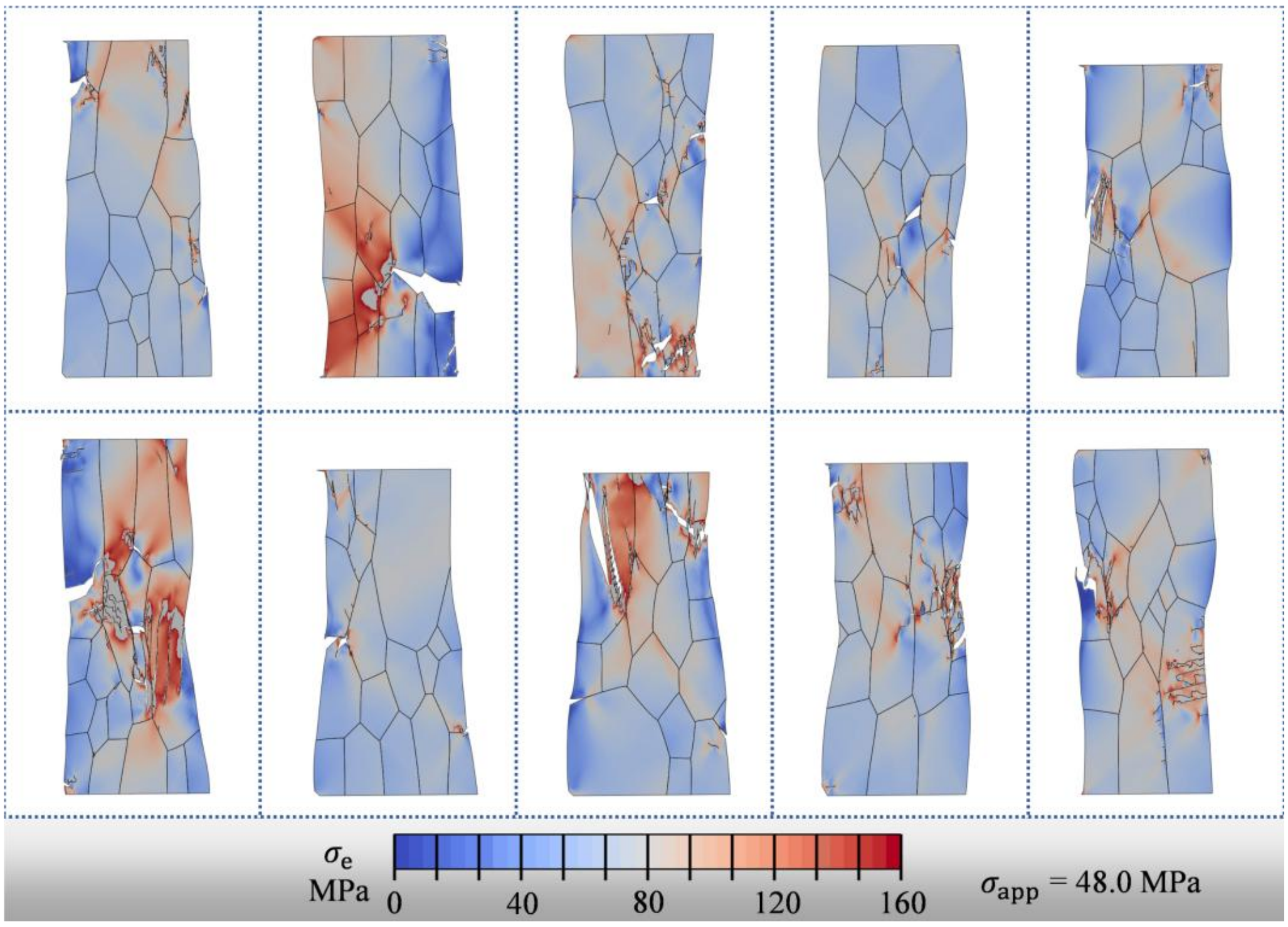

Fig. 18 Creep failure morphology of a random polycrystalline domains with a side length of 300 μm under uniaxial constant-stress creep loading at 48 MPa.

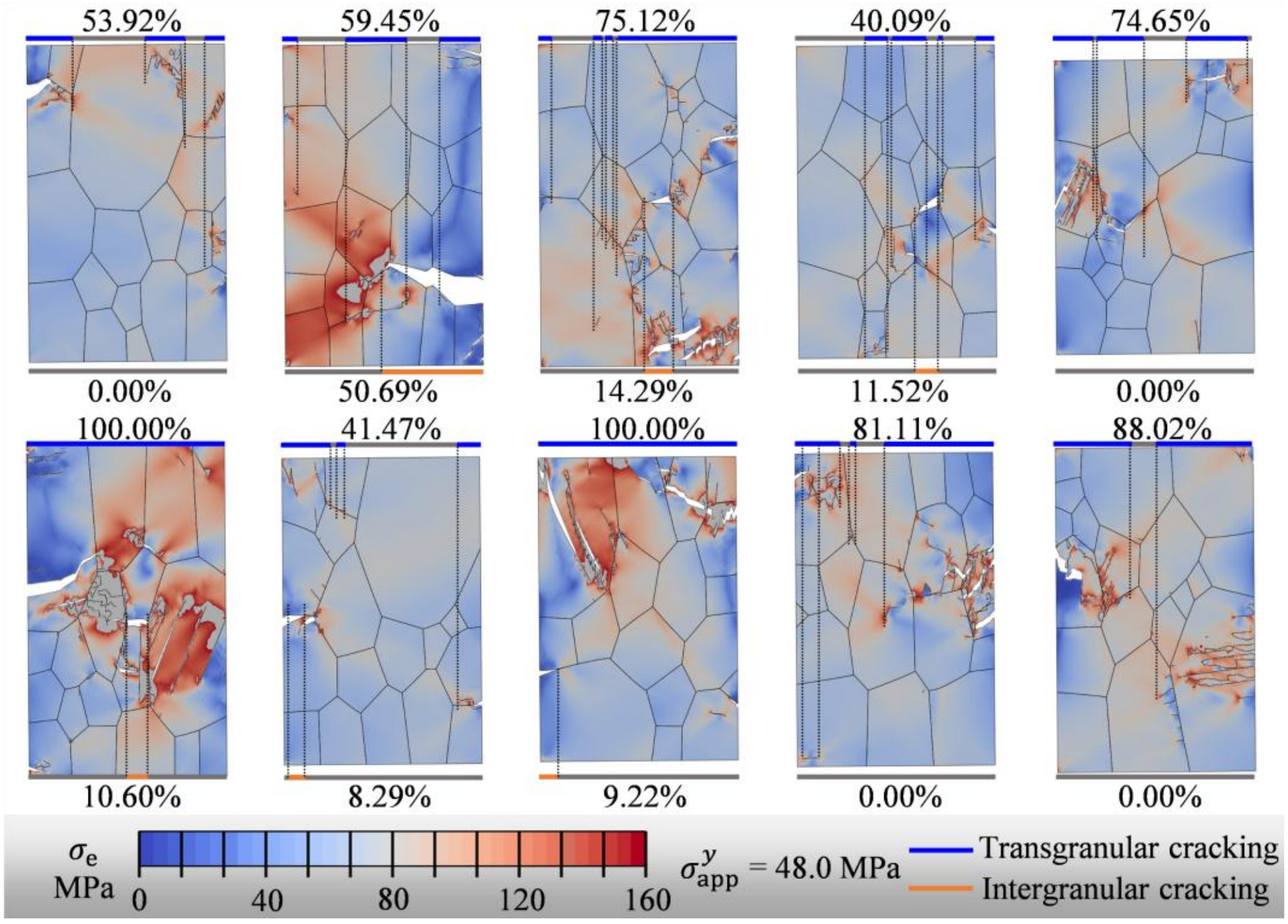

Fig. 19 Normalized lengths of inter-granular and trans-granular cracks with respect to the initial width of the polycrystalline domain, for a random polycrystalline domain with a side length of 300 μm under uniaxial constant-stress creep loading at 48 MPa.

To quantify this stress-sensitive transition in the cracking mode, the relative proportions of inter- and trans-granular cracks at failure are statistically evaluated for each polycrystalline domain. To ensure comparability among different polycrystalline domains, the crack length in each domain is normalized by its initial dimension perpendicular to the loading direction, which approximately corresponds to the dominant crack propagation direction. The corresponding statistical results for the five applied stress levels are presented in Fig. 11, 13, 15, 17, and 19. The projection of the crack length onto the direction perpendicular to the applied loading is defined as the effective crack length. The normalized effective crack lengths associated with inter- and trans-granular cracking are then statistically evaluated for each microstructural realization and averaged over the ten random polycrystalline domains at each applied stress level.

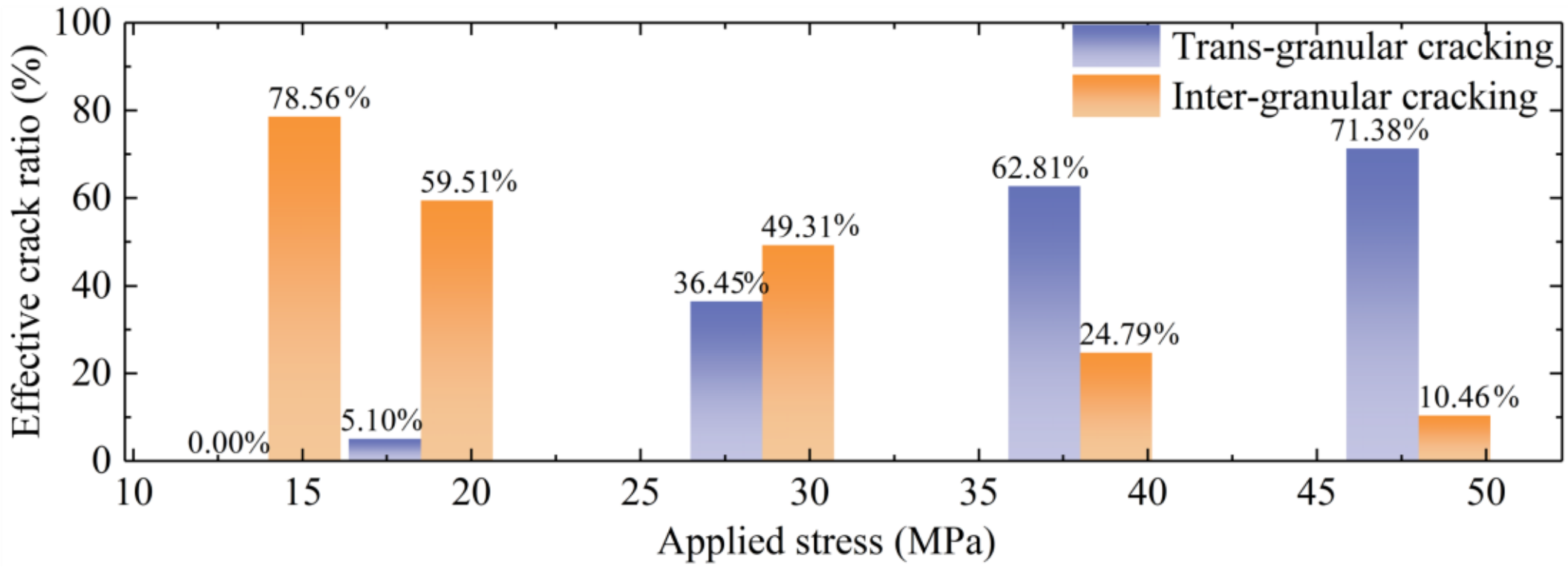

Fig. 20 Relative proportions of inter-granular and trans-granular cracking under different applied stress levels

The normalized effective crack ratios for inter-granular and trans-granular cracking are presented in Fig. 20, which quantitatively demonstrates the stress-dependent transition in the dominant cracking mode. As the applied stress increases, the normalized effective crack ratio associated with inter-granular cracking decreases progressively, whereas that associated with trans-granular cracking increases significantly. These results indicate that the proposed crystal plasticity-fracture framework captures the stress-dependent transition in creep cracking modes. The predicted transition is in good qualitative agreement with experimental observations (Natesan et al., 2012; Wang et al., 2021; Wang et al., 2022), where inter-granular fracture features coexist with trans-granular dimpled regions on the same creep fracture surface.

Furthermore, to illustrate the stress-dependent evolution of the two cracking modes, Fig. 21 shows the temporal evolution of the explicitly resolved trans-granular failure area fraction and inter-granular crack length fraction under four representative stress conditions. The explicitly developed trans-granular and inter-granular cracks are treated as two distinct forms of microscopic damage within the RVE. The trans-granular failure fraction is defined as the area fraction of failed grain elements:

$$F_{\mathrm{TG}} = \frac{A_{\mathrm{F}}}{A_{\mathrm{grain}}} \tag{43}$$

where $A_{\mathrm{F}}$ and $A_{\mathrm{grain}}$ denote the two-dimensional areas occupied by the failed trans-granular elements and by all grain elements, respectively. Similarly, the inter-granular crack failure fraction is defined as:

$$F_{\mathrm{IG}} = \frac{L_{\mathrm{F}}}{L_{\mathrm{GB}}} \tag{44}$$

where denote $L_{\mathrm{F}}$ and $L_{\mathrm{GB}}$ the total lengths of the fully failed GB elements and all GB elements, respectively. Both $F_{\mathrm{TG}}$ and $F_{\mathrm{IG}}$ are dimensionless quantities ranging from 0 to 1. However, because they represent different physical measures, a direct quantitative comparison between them is not appropriate. Nevertheless, their evolution with creep time provides a clear indication of the differences in cracking mechanisms under different stress levels. As shown in Fig. 21 a, at low stress, inter-granular failure develops markedly during the tertiary creep stage, whereas trans-granular cracking evolves only slowly. As shown in Fig. 21b, with increasing applied stress, inter-granular failure remains the dominant damage mode. However, during the final stage of overall failure, damage localization associated with inter-granular cracking leads to a rapid increase in trans-granular cracking in the vicinity of the localized damage region. This suggests that the initiation and accelerated development of trans-granular cracks are driven by the local strain elevation induced by damage localization resulting from inter-granular cracking. This behaviour is also evident in Fig. 7. As the applied stress increases to 38 MPa (Fig. 21c), trans-granular failure clearly becomes the dominant damage mode. Rather than exhibiting a sharp increase only at the final stage of rupture, the trans-granular failure fraction increases steadily throughout the tertiary creep stage. In contrast, the evolution of inter-granular failure remains relatively gradual. The widespread development of trans-granular cracking redistributes the damage within the microstructure and thereby suppresses the further localization and growth of inter-granular failure. Fig. 21d shows that, when the applied stress reaches 48 MPa, trans-granular failure becomes completely dominant, while inter-granular failure exhibits little further development.

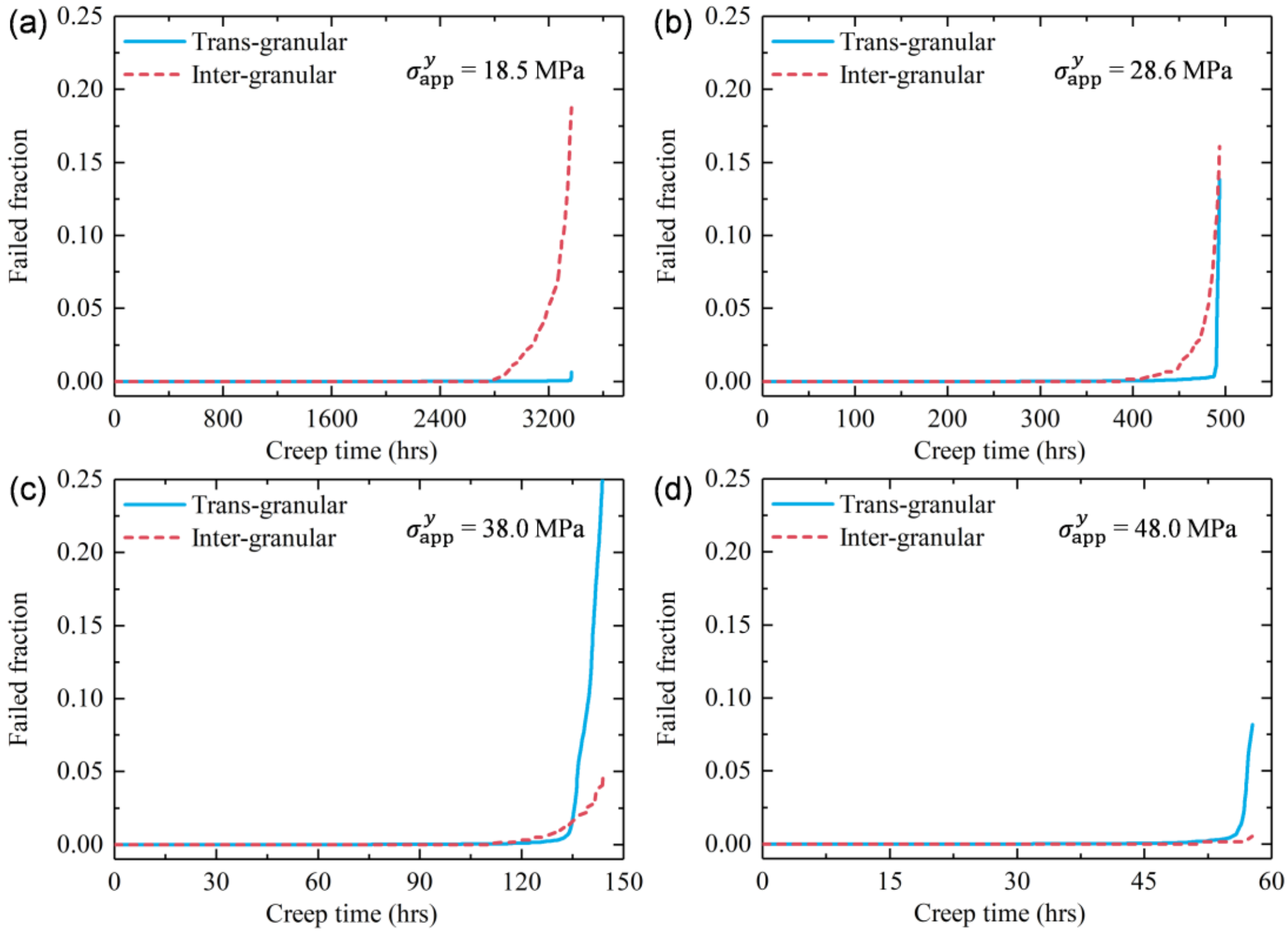

Fig. 21 Evolution of the area fraction of explicitly resolved trans-granular failure regions and the length fraction of explicitly captured inter-granular cracks at (a) 18.5 MPa, (b) 28.6 MPa, (c) 38 MPa, and (d) 48 MPa.

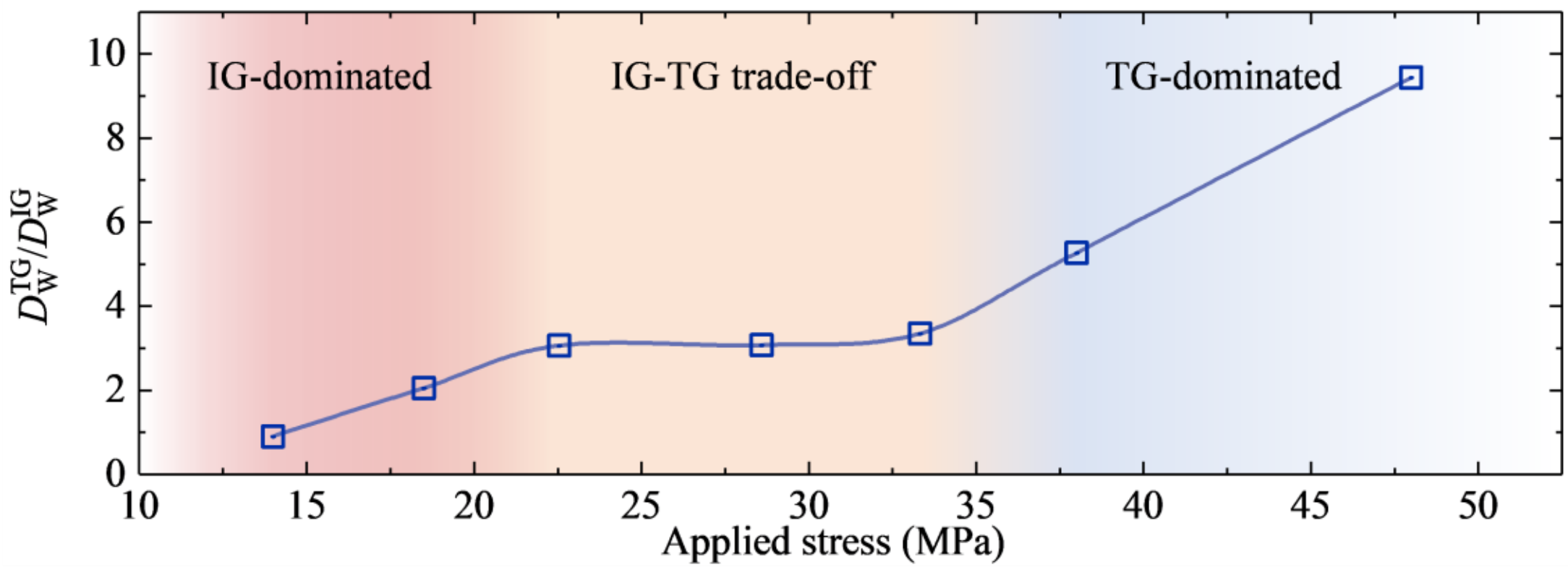

Fig. 22 Stress-sensitive proportions of equivalent trans-granular and inter-granular damage and the corresponding transition in the dominant fracture mechanism.

To provide a quantitative, RVE-averaged measure of the competition and trade-off between trans-granular and inter-granular damage from a global damage perspective, we define the global weighted trans-granular damage and weighted inter-granular damage as:

$$D_{\mathrm{W}}^{\mathrm{TG}} = \frac{\sum_i A_i \varGamma_i / \varGamma_{\mathrm{c}}}{A_{\mathrm{grain}}} \tag{45}$$

$$D_{\mathrm{W}}^{\mathrm{IG}} = \frac{\sum_i L_i D_i}{L_{\mathrm{GB}}} \tag{46}$$

where $A_i$ and $L_i$ denote the area of an individual grain element and the length of an individual GB element, respectively. $\varGamma_i$ and $D_i$ represent the corresponding slip measure and damage variable of element $i$, while $\varGamma_{\mathrm{c}}$ is the critical slip value defined in Eqn. (30). The ratio between these two quantities $D_{\mathrm{W}}^{\mathrm{TG}} / D_{\mathrm{W}}^{\mathrm{IG}}$ is then used to quantify the relative intensity of trans-granular and inter-granular damage. Fig. 22 shows that the relative contribution of trans-granular and inter-granular damage exhibits a clear stress dependence. At low stress, inter-granular damage dominates, whereas increasing stress leads to a progressive rise in the relative intensity of trans-granular damage. The plateau in the intermediate stress range reflects a dynamic trade-off between inter-granular and trans-granular damage. Although increasing stress continuously shifts the fracture mechanism toward trans-granular cracking, the two damage modes evolve in a compensating manner: inter-granular localization promotes trans-granular cracking, while the spread of trans-granular damage redistributes the local stress field and suppresses further inter-granular localization. Their relative damage intensities therefore remain nearly balanced, producing the observed plateau. This plateau therefore represents a genuine competition regime rather than a cessation of damage evolution.

### 4.3. Damage localization associated with the creep cracking transition

As demonstrated in Sections 4.1 and 4.2, the transition of creep failure mechanisms originates from the competition and interaction between time-controlled GB damage accumulation and deformation-controlled trans-granular fracture processes. At low stress levels, prolonged creep exposure promotes progressive GB damage accumulation, which eventually leads to localized inter-granular fracture. At high stress levels, substantial creep deformation accumulation induces intragranular damage localization and trans-granular cracking before pronounced GB degradation develops. At intermediate stress levels, a complex interaction between inter-granular and trans-granular cracking occurs. On the one hand, the development of one cracking mode can promote the initiation and growth of the other damage mode; on the other hand, the coexistence of multiple cracking modes leads to competition and mutual suppression. Corresponding to the trade-off regime in Fig. 22, the relative contribution of trans-granular damage does not continuously increase with increasing applied stress, but instead remains relatively stable within a certain range. This behaviour indicates the existence of a dynamic competition and compensating interaction between inter-granular and trans-granular damage mechanisms.

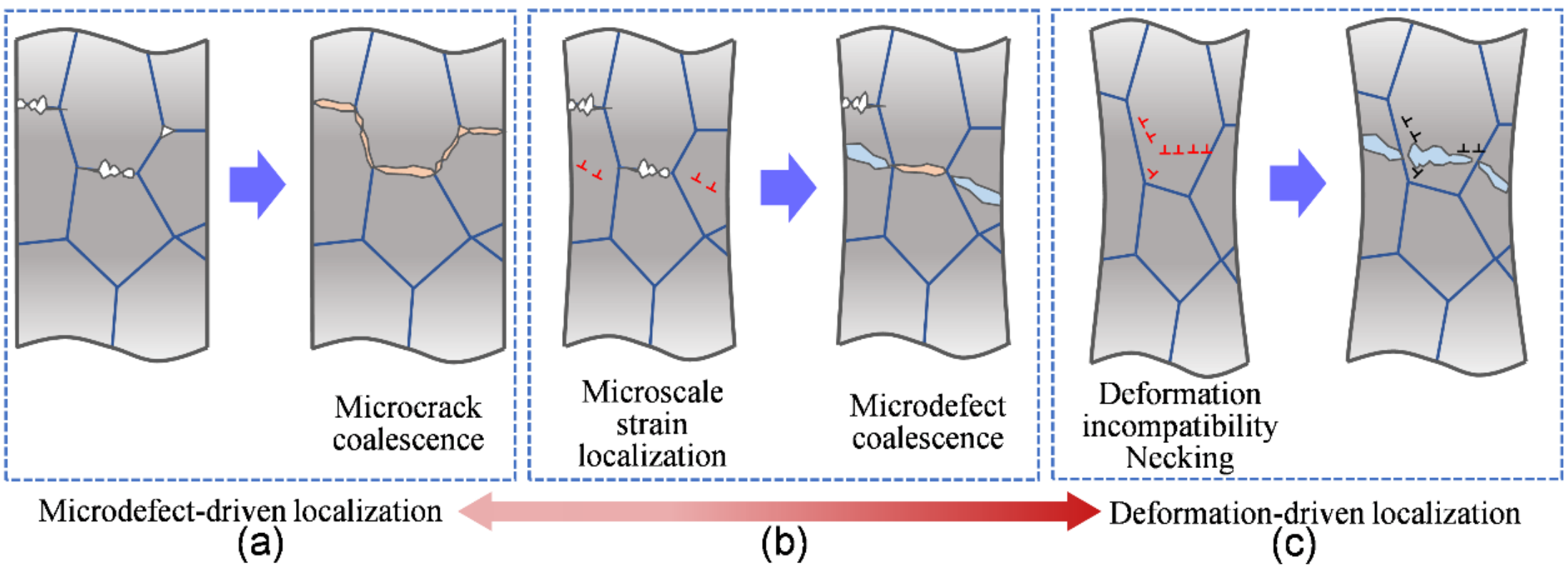

Fig. 23 Schematic illustration of distinct creep rupture localization mechanisms associated with time-controlled GB degradation and deformation-driven intragranular damage.

From the perspective of damage evolution, the transition and competition between these two mechanisms give rise to three distinct modes of damage localization. As shown in Fig. 23a, at low stress levels, corresponding to the regime dominated by time-dependent GB damage, GB degradation is initially distributed relatively uniformly throughout the material, and no significant macroscopic damage localization occurs at this stage. Damage localization emerges only during the final stage of failure, driven by the coalescence of dispersed microscopic creep cracks. This behaviour is clearly observed in Fig. 6, where the final failure exhibits a macroscopically quasi-brittle fracture response (Fig. 9a). At high stress levels, Fig. 23c demonstrates a fundamentally different localization mechanism. Before pronounced intragranular creep damage develops, deformation incompatibility arising from the heterogeneous nature of the polycrystalline structure already determines the preferential regions for damage localization. With increasing creep exposure, localized necking and trans-granular fracture subsequently develop within these regions. Therefore, the localization process in this regime is primarily associated with deformation incompatibility-induced localization. This process is illustrated in Fig. 8 and leads to the macroscopic necking localization shown in Fig. 9c. As shown in Fig. 23b, the damage localization behaviour in the intermediate stress regime is more complex, reflecting the characteristic trade-off between different damage mechanisms. During the early stage of creep damage accumulation, progressive GB damage induces microscale stress concentration, which further promotes strain localization in the grains adjacent to the damaged GBs and subsequently triggers trans-granular cracking. Meanwhile, the development of trans-granular cracks can also influence the subsequent evolution of local GB damage accumulation. At the later stage of creep failure, the interaction between these damage processes ultimately leads to the coalescence of multiple microscopic defects, including both trans-granular and inter-granular cracks. This process is illustrated in Fig. 7. As shown in Fig. 9b, it also leads to a macroscopically quasi-brittle fracture response.

Therefore, with increasing applied creep stress, the damage localization mode gradually transitions from microdefect-driven localization to deformation-driven localization. The intermediate stress regime represents a mixed localization regime, in which microdefect-driven and deformation-driven localization mechanisms coexist and interact, giving rise to the observed trade-off between inter-granular and trans-granular damage.

## 4.4. Mechanism-dependent macroscopic rupture ductility

Moreover, the transition in creep cracking mechanism has a pronounced influence on the macroscopic creep response of the material, particularly on the creep rupture strain (see Fig. 3). As summarized by Wen et al. (2016), the variation of creep rupture strain with applied stress is a common feature of many metallic materials, including alloys and stainless steels. Based on the responses simulated by the crystal plasticity-fracture framework, the macroscopic creep strain curves of the random polycrystalline configurations are obtained at different applied stress levels and shown in Fig. 24. As shown in Fig. 24a, the creep rupture strains at 38 and 48 MPa remain within a similar range (approximately 0.5~0.7), despite the substantial difference in creep life. In contrast, Fig. 24b~d show that the macroscopic creep rupture strain decreases markedly as the applied stress is reduced.

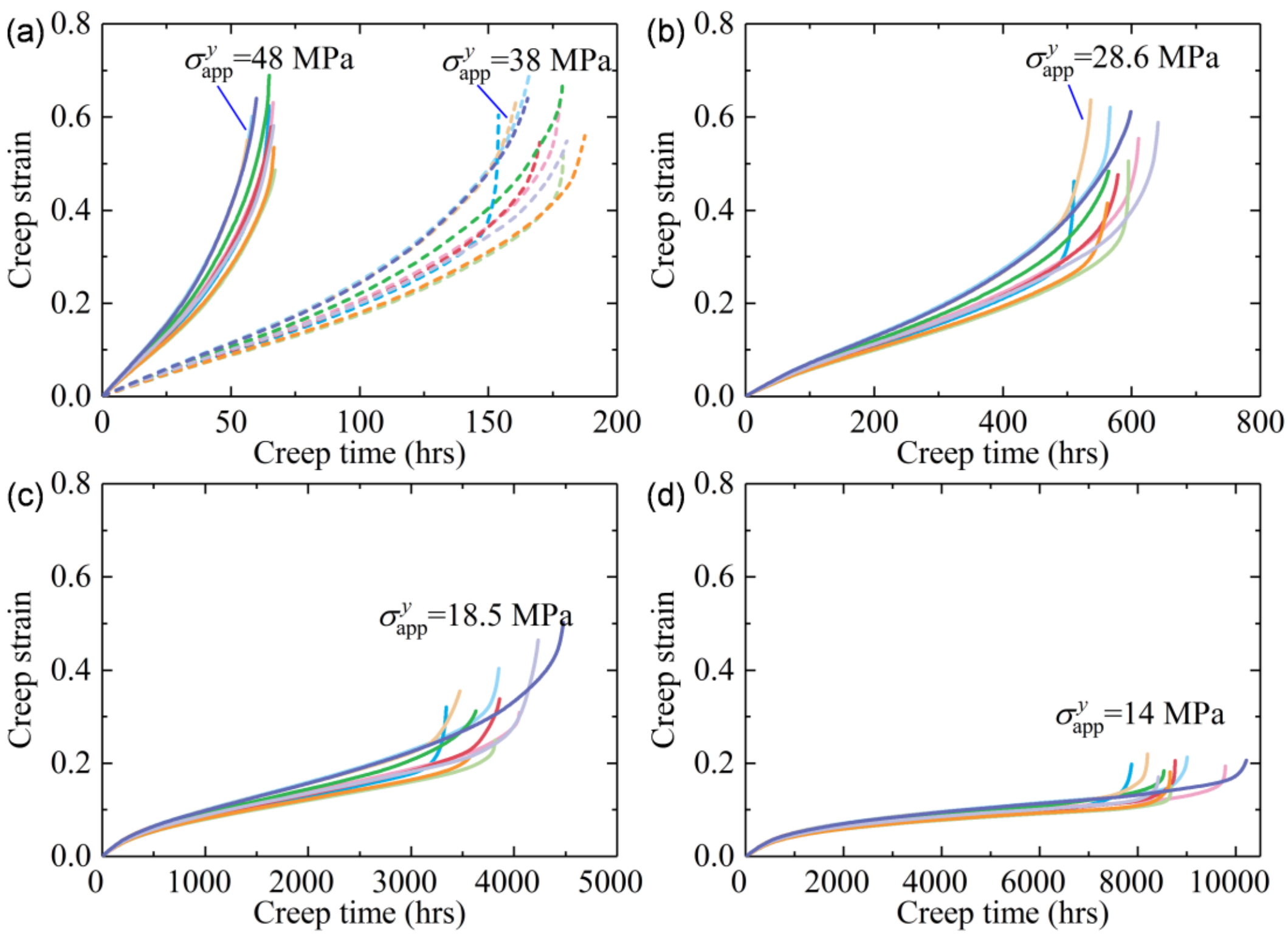

Fig. 24 The macroscopic creep strain curves of the random polycrystalline configurations under different applied stress: (a) 38 and 48 MPa; (b) 28.6 MPa; (c) 18.5 MPa; (d) 14 MPa

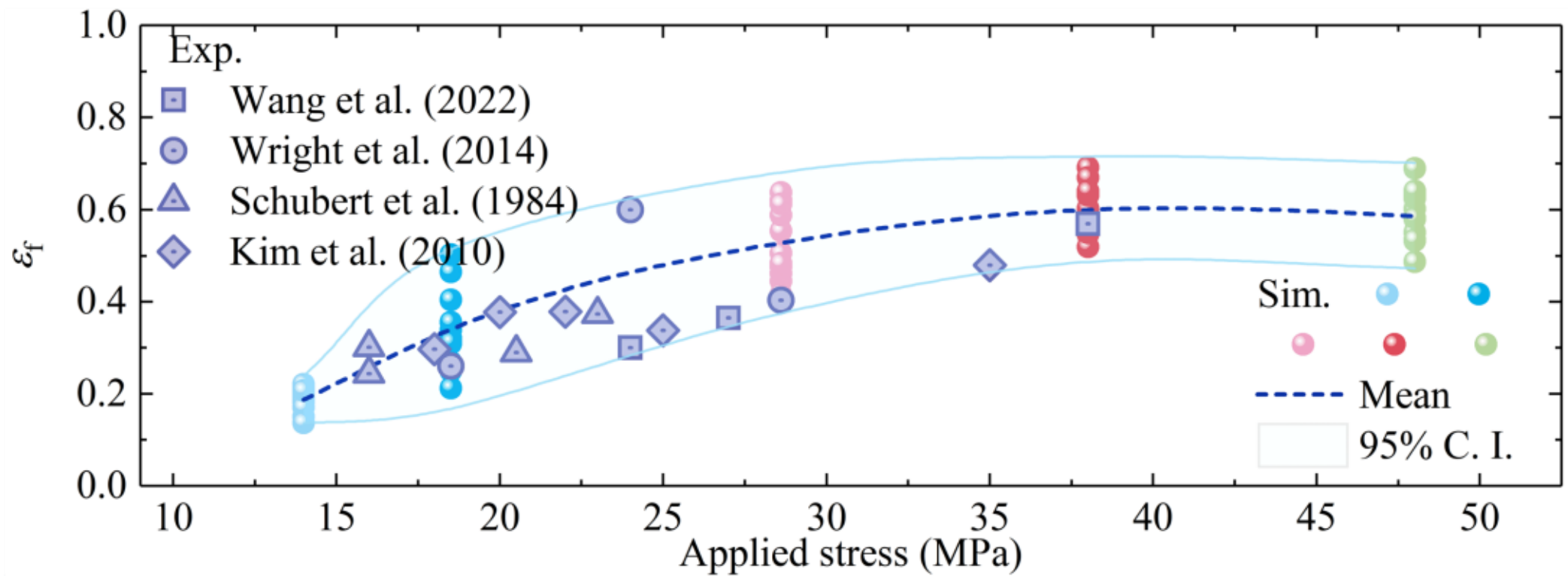

Fig. 25 Dependence of creep rupture strain on the applied stress: comparison between simulations and experimental data

The creep rupture strains at different applied stress levels are presented in Fig. 25, together with the mean values and 95% confidence intervals (C.I.) obtained from the random polycrystalline realizations. With increasing applied stress, the creep rupture strain increases progressively and eventually approaches a plateau. The micromechanical analyses in Sections 4.2~4.4 explain the origin of this stress dependence. At high stress levels, trans-granular cracking dominates the failure process, and rupture is primarily associated with the accumulation of macroscopic deformation, giving rise to a strain-controlled failure response. At low stress levels, failure is instead dominated by inter-granular cracking and exhibits a quasi-brittle character. In this regime, the rupture process is controlled to a greater extent by the progressive, time-controlled degradation of the GBs, leading to a predominantly time-controlled creep failure response. For Inconel 617 at 950 °C, the present results suggest that the transition from quasi-brittle to ductile creep failure occurs approximately within the applied stress range of 15~35 MPa. Fig. 25 also demonstrates good agreement between the predicted creep rupture strains and the experimental data (Kim et al., 2010; Schubert et al., 1984; Wang et al., 2022; Wright et al., 2014), providing further validation of the proposed crystal plasticity-fracture framework for creep failure analysis of Inconel 617.

Moreover, comparison between Fig. 25 and Fig. 22 shows that the macroscopic transition from low to high creep rupture ductility largely coincides with the microscopic IG-TG trade-off regime. This correspondence can be attributed to the change in damage localization associated with the IG-TG trade-off: the increasing participation of trans-granular cracking redistributes damage away from highly localized GB paths, thereby allowing greater macroscopic deformation before final rupture. As the stress increases further, trans-granular cracking progressively becomes dominant, accompanied by greater participation of intragranular plastic deformation and, consequently, higher macroscopic creep rupture ductility.

## 4.5. Creep life assessment considering the mechanism transition

Similarly, the uniaxial creep rupture time at different applied stress levels is evaluated based on the simulation results. Fig. 26 presents the predicted and experimental uniaxial creep rupture times of Inconel 617 at 950 °C (Kim et al., 2010; Schubert et al., 1984; Wang et al., 2022; Wright et al., 2014). Overall, the predicted results show good agreement with the experimental data over a relatively wide range of applied stress levels. Moreover, the proposed crystal plasticity–fracture framework successfully captures the transition in the creep rupture time curve at low stress levels. Specifically, the proposed model accurately captures the change in the slope of the low-stress region of the creep rupture time curve in the log-log plot. Such a transition in the creep rupture time is commonly observed in high-temperature creep of metallic materials and has been widely reported in the literature (Phan et al., 2017; Sanders et al., 2017; Zhang et al., 2020). This transition is generally considered to arise from a change in the dominant GB cavity growth mechanism, from bulk-creep-controlled growth (see zone 'C' in Fig. 26) at relatively high stresses to diffusion-controlled growth (see zone 'D' in Fig. 26) at lower stresses. In other words, the transition discussed above is generally interpreted in terms of the dominant GB cavity growth mechanism, whereas the present work focuses on the stress-dependent transition in the dominant cracking mode between inter-granular and trans-granular fracture. Although both phenomena occur during creep failure, they describe different aspects of the underlying failure process. In the present CP–fracture framework, the bulk-creep-controlled and diffusion-controlled GB cavity growth mechanisms are phenomenologically distinguished through the formulations in Eqns.(24)~(26). Consequently, the change in their relative contributions with stress can be reflected in the predicted creep rupture response, allowing the model to reproduce, to some extent, the low-stress transition in the creep rupture time curve.

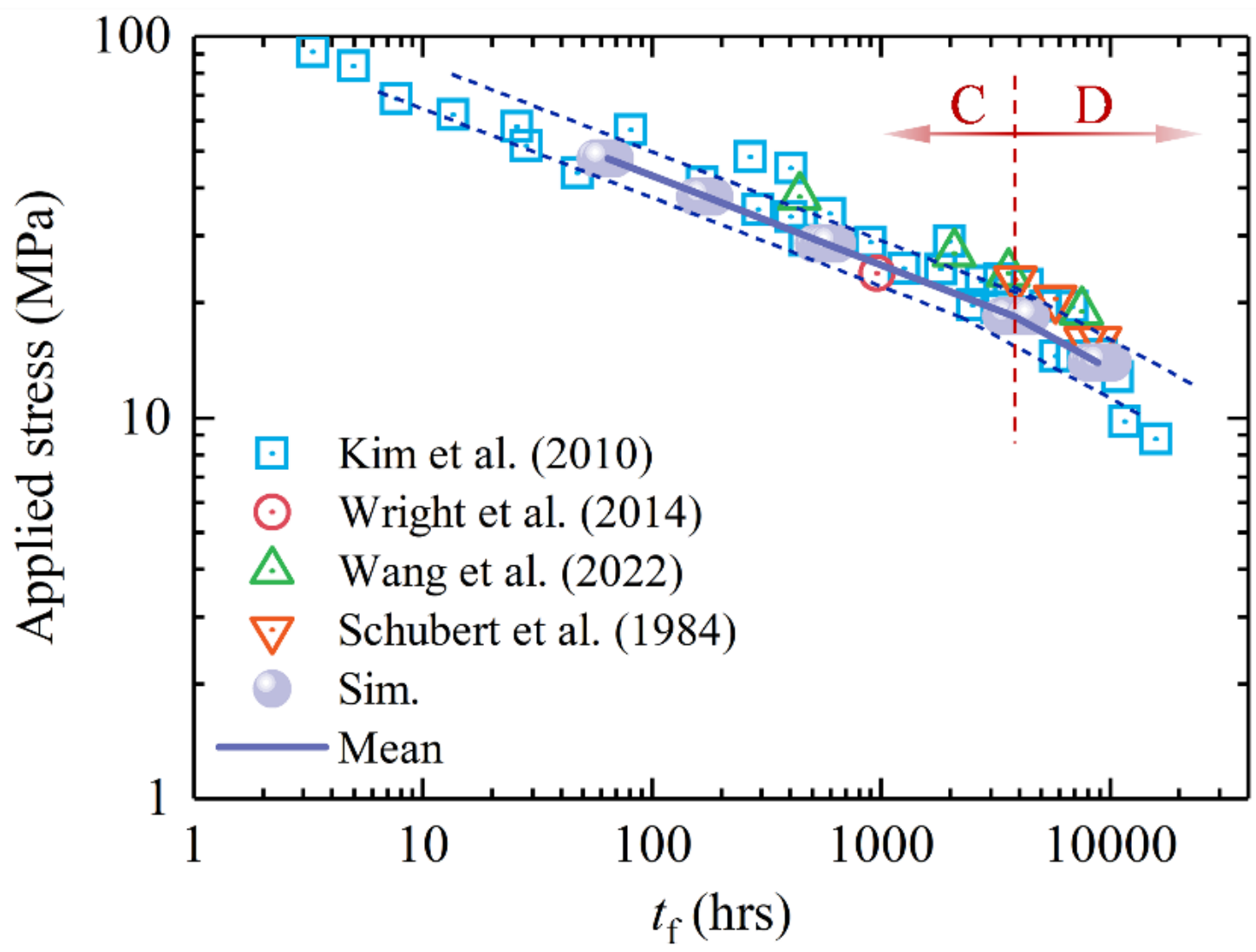

Fig. 26 Uniaxial creep rupture lifetime of Inconel 617 at 950 °C under different applied stress levels.

### 4.6. Creep cracking under the biaxial stress state

Based on the proposed crystal plasticity-fracture framework, creep behaviour under multiaxial loading conditions can also be simulated, allowing the corresponding creep rupture time to be predicted. The present simulations focus on Inconel 617 at 950 °C and are validated against experimental data from internally pressurized tubes (Wright and wright, 2013), which provide a representative multiaxial stress state for high-temperature piping applications. To represent the biaxial stress state of internally pressurized tubes, the loading condition in Fig. 5a is prescribed with a stress ratio of $\sigma^{y}_{\mathrm{app}} / \sigma^{x}_{\mathrm{app}} = 2$.

As shown in Fig. 27, the predicted results agree well with the experimental data. This agreement supports the capability of the proposed micromechanical framework to describe and predict creep damage and failure under multiaxial stress states. Fig. 28 presents the creep damage and failure behaviour under biaxial stress states. As shown in Fig. 28 a, c and d, inter-granular cracking constitutes the dominant failure mode in the internally pressurized tube. The major cracks propagate approximately perpendicular to the direction of the macroscopic maximum principal stress, while the overall macroscopic response exhibits quasi-brittle fracture characteristics. As shown in Fig. 28b, a small amount of localized trans-granular cracking develops near the tips of propagating inter-granular cracks, accompanying the advance of the inter-granular cracks. This behaviour differs from that observed under uniaxial loading. At a comparable equivalent stress level, pronounced trans-granular cracking is already observed in the uniaxial case, while the biaxial case remains predominantly characterized by inter-granular cracking. This difference is associated with the hydrostatic stress component introduced by the biaxial stress state. From the macroscopic stress perspective, the von Mises equivalent stress, which governs slip-dominated plastic deformation, reaches a relatively high level. Nevertheless, GB degradation is sensitive not only to the equivalent stress but also to the tensile principal stress and hydrostatic stress components. The elevated hydrostatic stress under biaxial loading accelerates GB degradation and consequently promotes inter-granular cracking, causing it to remain the dominant failure mode.

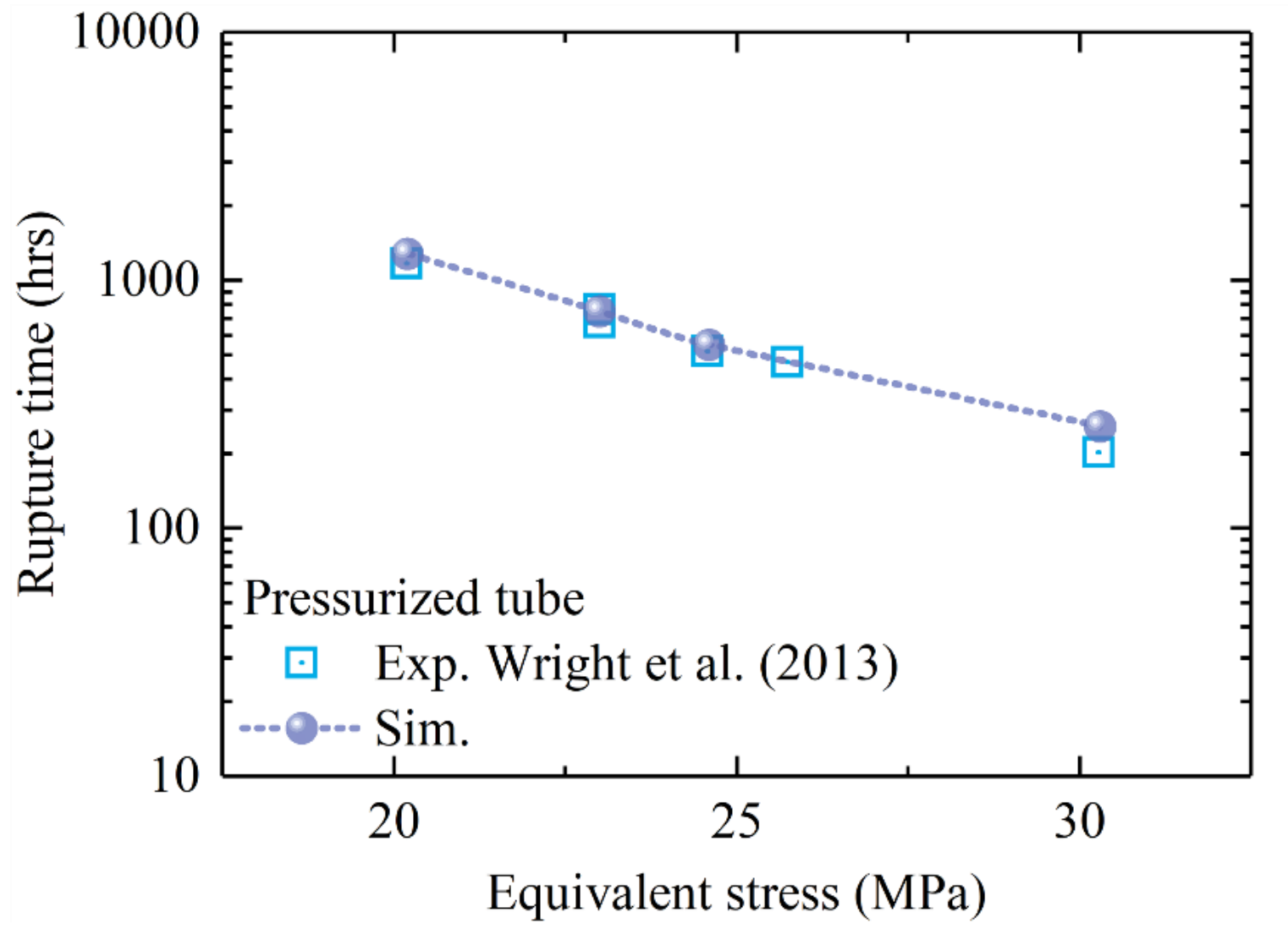

Fig. 27 Predicted and experimental creep rupture lifetimes of internally pressurized Inconel 617 tubes under biaxial stress conditions.

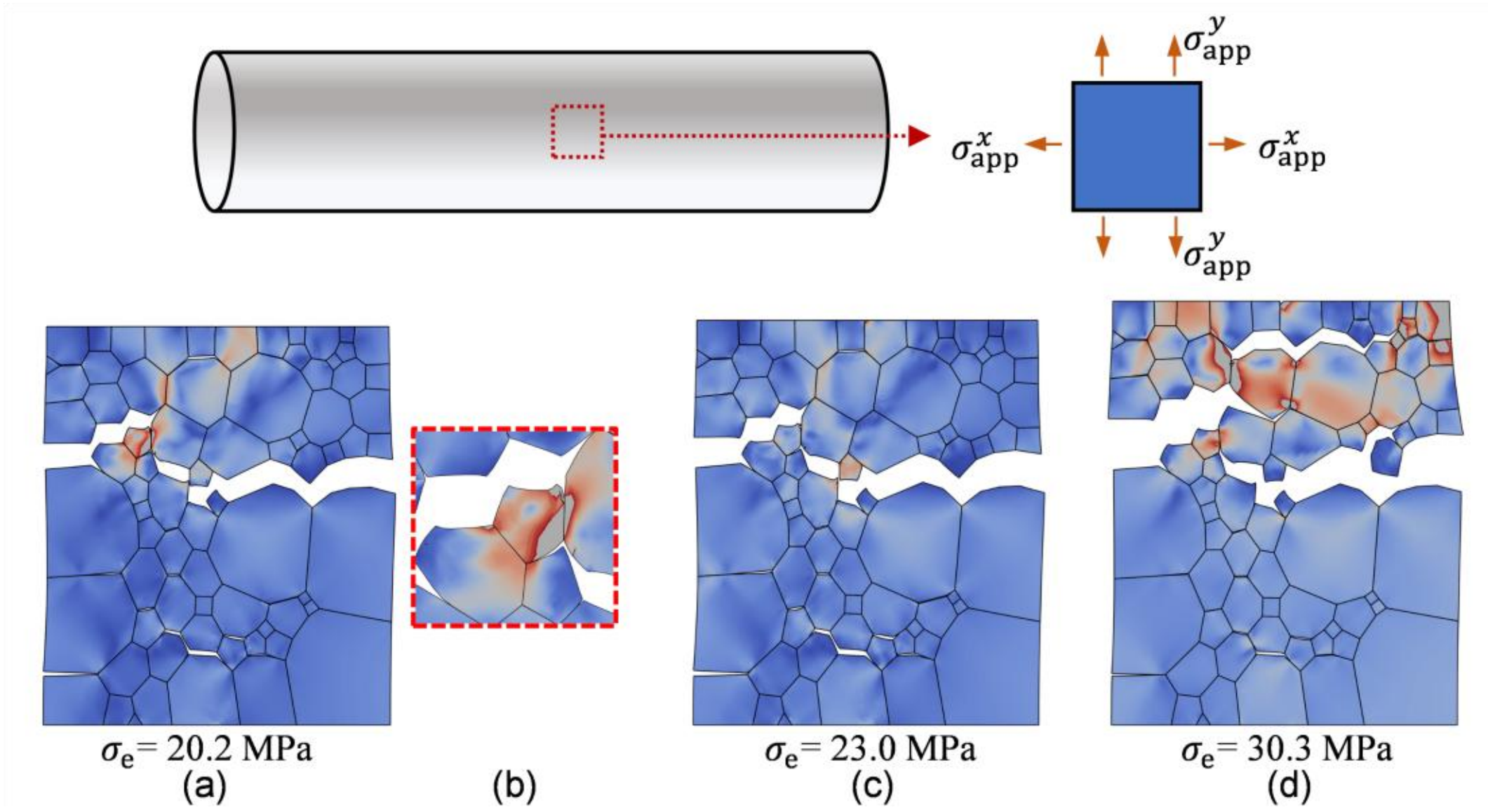

Fig. 28 Simulated creep rupture morphologies of internally pressurized Inconel 617 tubes under biaxial loading at different equivalent stress levels: (a) 20.2 MPa; (b) enlarged view showing the local crack morphology; (c) 23.0 MPa; (d) 30.3 MPa.

# 5. Usability and potential extension

The creep deformation and fracture behaviour of metallic materials is governed by multiple interacting mechanisms, whose relative importance depends on both the applied stress and temperature. According to the classical creep fracture mechanism map proposed by Ashby et al. (1979), different failure modes may dominate under different loading conditions, including plasticity-controlled ductile fracture, trans-granular creep fracture, inter-granular creep fracture, and creep rupture associated with dynamic recrystallization at very high temperatures. Developing a unified framework capable of explicitly resolving all these mechanisms and their transitions within a single model remains highly challenging. The present work therefore focuses on the transition between inter-granular and trans-granular creep cracking under typical high-temperature service conditions. A crystal plasticity-fracture framework is developed to capture the competition between these two dominant cracking modes and to establish the relationship between the underlying micromechanical processes and the corresponding macroscopic creep response, including crack evolution, creep rupture strain, and creep rupture lifetime.

The objective of the present framework is therefore not to reproduce the entire creep fracture mechanism map, but rather to provide a physically consistent mesoscale description of the competition between inter-granular and trans-granular cracking within the stress and temperature regime where these two mechanisms dominate. In addition, the GB degradation formulation phenomenologically distinguishes between bulk-creep-controlled and diffusion-controlled cavity growth mechanisms. Although vacancy transport and diffusion kinetics are not explicitly resolved, this formulation allows the dominant contribution to GB cavity growth to vary with the loading condition, thereby providing a physically interpretable description of GB degradation and its influence on creep rupture. Although the present study focuses on Inconel 617, the transition between inter-granular and trans-granular creep fracture is widely observed in a broad range of metallic materials. Therefore, the proposed framework is expected to retain a certain degree of transferability across different material systems, provided that the corresponding constitutive and fracture parameters are recalibrated for the material of interest. The proposed framework also demonstrates applicability under multiaxial stress states. This capability provides a useful basis for constructing fracture mechanism maps under multiaxial loading conditions.

For Inconel 617 at 950 °C, however, additional microstructural processes may also influence creep failure. Experimental observations reported by Wang et al. (2021)

indicate that, in addition to the conventional GB degradation associated with creep damage, dynamic recrystallization can become significant under high-temperature and high-stress conditions. At relatively lower temperatures and stress levels, long-term carbide precipitation and coarsening may also contribute to material embrittlement. These mechanisms are not explicitly represented in the present framework and may affect its quantitative applicability outside the stress-temperature regime dominated by the transition between inter-granular and trans-granular cracking. These additional mechanisms provide natural directions for further extension of the present framework. For instance, carbide-induced embrittlement can be incorporated by introducing an intragranular cleavage-fracture formulation, thereby extending the model to creep conditions where brittle trans-granular fracture becomes significant.

# 6. Conclusions

A crystal plasticity-fracture framework is established to unravel the stress-dependent transition from inter-granular to trans-granular creep cracking in Inconel 617 at 950 °C. By coupling crystal deformation, GB degradation, and trans-granular fracture within a unified finite element formulation, the framework enables the evolution of microscale damage mechanisms and their quantitative linkage with macroscopic creep rupture behaviour to be systematically investigated. The main conclusions are summarized as follows.

(1) The proposed unified crystal plasticity-fracture framework successfully captures the coupled evolution of creep deformation, GB degradation, and inter- and trans-granular cracking within explicit polycrystalline microstructures. The predicted macroscopic creep strain response, rupture strain, and creep life show good agreement with the available experimental data over a broad range of applied stress levels, providing validation for investigating the underlying mechanisms governing creep fracture transition in Inconel 617.

(2) The stress-dependent transition from inter-granular to trans-granular creep fracture is quantitatively identified and mechanistically explained. At relatively low stresses, creep damage is dominated by progressive GB degradation and inter-granular cracking, whereas trans-granular cracking becomes increasingly significant with increasing stress and eventually dominates at high stresses. This shift in the relative contribution of different damage mechanisms is accompanied by a transition in damage localization behaviour, leading to the evolution from inter-granular-dominated to mixed and finally trans-granular-dominated fracture. For Inconel 617 at 950 °C, the transition from quasi-brittle, inter-granular failure to ductile, trans-granular failure occurs approximately within the applied stress range of 15~35 MPa.

(3) A micromechanical trade-off between inter-granular and trans-granular cracking is revealed, establishing its linkage with the macroscopic creep rupture ductility transition. The competition between these two cracking mechanisms leads to distinct damage localization modes under different stress regimes. At low stress levels, time-dependent GB damage accumulation causes microcrack coalescence-induced localization and localized inter-granular fracture, whereas high stresses promote deformation incompatibility-driven strain localization and spatially distributed trans-granular damage development. During the transition regime, inter- and trans-granular damage interact competitively: inter-granular localization induces local stress concentration and facilitates trans-granular cracking, while the subsequent development of trans-granular damage redistributes the local damage field and limits further inter-granular localization. This coupled evolution results in the coalescence of multiple microscopic defects and a balanced contribution of the two damage modes. As trans-granular cracking becomes dominant, damage distribution becomes more dispersed, enabling greater macroscopic deformation before rupture and enhancing creep rupture ductility.

(4) The effect of multiaxial stress state on the competition between inter-granular and trans-granular cracking is clarified. Under biaxial loading representative of internally pressurized tubes, elevated hydrostatic and principal stresses promote GB degradation and inter-granular cracking, even at equivalent stress levels where pronounced trans-granular cracking develops under uniaxial loading. The predicted creep lives under biaxial loading agree well with the corresponding experimental results, demonstrating the applicability of the framework to multiaxial creep failure analysis.

Overall, the present framework provides a physically interpretable mesoscale description of the transition between inter-granular and trans-granular creep fracture and its influence on macroscopic rupture behaviour. Although the current formulation focuses on Inconel 617 and does not explicitly resolve all high-temperature degradation mechanisms, its modular structure provides a basis for further extensions involving three-dimensional microstructures, explicit diffusion-damage coupling, additional brittle or recrystallization-related fracture mechanisms, and the construction of multiaxial creep fracture mechanism maps.

## Acknowledgments

The study was supported by the Key Project (12332005) and the Youth Science Fund (12602118) of the National Natural Science Foundation of China, and the Postdoctoral Fellowship Program of CPSF under Grant Number GZC20261057.